\documentclass[10pt,twocolumn,headsepline,DIV=22,paper=A4]{scrartcl}

\pdfoutput=1

\usepackage[UKenglish]{babel}
\usepackage[UKenglish,cleanlook]{isodate}
\usepackage[T1]{fontenc}
\usepackage{amsmath}
\usepackage{amssymb}
\usepackage{amsthm}
\usepackage{IEEEtrantools}
\usepackage[bb=boondox]{mathalfa}
\usepackage{stmaryrd}
\usepackage{cases}
\usepackage{mathtools}
\usepackage{tabu,multirow,colortbl}
\usepackage[table]{xcolor}
\usepackage{array}
\usepackage{titling}
\usepackage{authblk}
\usepackage{abstract}
\usepackage{enumitem}
\usepackage{microtype}
\usepackage{verbatim}
\usepackage[square,comma,numbers,sort&compress]{natbib}
\usepackage[colorlinks,linkcolor=blue,citecolor=blue,urlcolor=blue]{hyperref}
\usepackage{quant_defs}

\pretitle{\centering\large\normalfont\bfseries}
\posttitle{\par\vskip 0.5cm}
\preauthor{\centering}
\postauthor{\par}
\predate{\centering\small(Dated: }
\postdate{)\par}

\AtBeginDocument{}

\setcapindent{1em}

\setkomafont{disposition}{\rmfamily\bfseries\boldmath\small}
\setkomafont{section}{\centering\MakeUppercase}
\setkomafont{subsection}{\centering}
\setkomafont{subsubsection}{\centering\itshape\textmd}
\setkomafont{paragraph}{\normalsize\itshape\textmd}

\newcommand{\email}[1]{\thanks{\href{mailto:#1}{#1}}}

\newcommand{\mkTitleAndAbstract}[1]{%
  \twocolumn[
    \maketitle
    \begin{onecolabstract}
      #1 \vspace{0.5cm}
    \end{onecolabstract}
  ]
  \saythanks}

\def\amjphys{Am.\ J.\ Phys.}%
\def\jphysa{J.\ Phys.\ A: Math.\ Theor.}%
\def\jphysagen{J.\ Phys.\ A: Math.\ Gen.}%
\def\lmp{Lett.\ Math.\ Phys.}%
\def\nature{Nature}%
\def\physics{Physics}%
\def\pra{Phys.\ Rev.\ A}%
\def\prl{Phys.\ Rev.\ Lett.}%
\def\quantic{Quantum Inf.\ Comput.}%
\def\rmp{Rev.\ Mod.\ Phys.}%

\newtheorem{lemma}{Lemma}

\newcommand{\vect}[1]{\boldsymbol{#1}}

\newcommand{\rA}{\mathrm{A}}
\newcommand{\rB}{\mathrm{B}}

\newcommand{\rABE}{\mathrm{ABE}}
\newcommand{\rE}{\mathrm{E}}
\newcommand{\rx}{\mathrm{x}}
\newcommand{\ry}{\mathrm{y}}
\newcommand{\rz}{\mathrm{z}}
\newcommand{\sx}{\mathrm{X}}
\newcommand{\sy}{\mathrm{Y}}
\newcommand{\sz}{\mathrm{Z}}
\newcommand{\sv}{\vect{\sigma}}

\newcommand{\hilb}{\mathcal{H}}

\DeclareMathOperator{\re}{Re}

\newcolumntype{P}[1]{>{\centering\arraybackslash}p{#1}}

\newcommand{\liq}{Laboratoire d'Information Quantique, CP~225, Universit\'{e}
  libre de Bruxelles, \newline Avenue F.~D.~Roosevelt 50, 1050 Bruxelles,
  Belgium}

\newcommand{\uow}{Faculty of Physics, University of Warsaw, Pasteura 5,
  02-093 Warsaw, Poland}

\newcommand{\icfo}{ICFO -- Institut de Ciencies Fotoniques, The
  Barcelona Institute of Science and Technology, \newline
  08860 Castelldefels (Barcelona), Spain}

\newcommand{\icrea}{ICREA -- Institucio Catalana de Recerca i Estudis
  Avan\c{c}ats, Lluis Companys 23, 08010 Barcelona, Spain}

\newcommand{\ctp}{Center for Theoretical Physics, Polish Academy of Sciences,
  Aleja Lotnik\'o{}w 32/46, 02-668 Warsaw, Poland}

\newcommand{\rsection}[1]{\textit{#1.---}}

\title{Maximal randomness from partially entangled states}

\author[1,2]{Erik~Woodhead\email{erik.woodhead@ulb.ac.be}}
\author[3,4]{J\k{e}drzej~Kaniewski\email{jkaniewski@fuw.edu.pl}}
\author[2]{Boris~Bourdoncle}
\author[2]{Alexia~Salavrakos}
\author[2]{Joseph~Bowles}
\author[2,5]{Antonio~Ac\'{\i}n}
\author[4]{Remigiusz~Augusiak}

\affil[1]{\liq}
\affil[2]{\icfo}
\affil[3]{\uow}
\affil[4]{\ctp}
\affil[5]{\icrea}

\date{13~November~2020}

\begin{document}

\mkTitleAndAbstract{%
  We investigate how much randomness can be extracted from a generic
  partially entangled pure state of two qubits in a device-independent
  setting, where a Bell test is used to certify the correct functioning of
  the apparatus. For any such state, we first show that two bits of
  randomness are always attainable both if projective measurements are used
  to generate the randomness globally or if a nonprojective measurement is
  used to generate the randomness locally. We then prove that the maximum
  amount of randomness that can be generated using nonprojective measurements
  globally is restricted to between approximately 3.58 and 3.96 bits. The
  upper limit rules out that a bound of four bits potentially obtainable with
  extremal qubit measurements can be attained. We point out this is a
  consequence of the fact that nonprojective qubit measurements with four
  outcomes can only be self-tested to a limited degree in a Bell experiment.
}

Although it was not the original motivation \cite{ref:n2011}, Bell's
theorem~\cite{ref:b1964} allows for a very strong test of quantum
randomness. By preparing an entangled quantum system and exhibiting a Bell
inequality violation with it, we can immediately know that the measurement
outcomes were not the result of an underlying deterministic process. This
observation is the basis of a class of quantum cryptography protocols, called
\emph{device independent}, that incorporate a Bell test as a self-test of the
correct functioning of the apparatus. The class includes device-independent
versions of quantum key distribution and random number
generation~\cite{ref:ab2007,ref:c2006,ref:pam2010,ref:bc2014}.

This perspective prompts an obvious question: How much randomness can we
extract from a given quantum system, and how might this depend on the degree
of entanglement? Previous work (see table~\ref{tab:context}) has indicated
that the two do not seem strongly related; we cannot necessarily get more
randomness from a maximally entangled state than a weakly entangled one of
the same dimension. This point was first made in \cite{ref:amp2012} where it
was shown that, with the help of a suitable Bell test, a uniformly random bit
could be generated from the result of a projective measurement performed on
one part of any partially entangled pure state of two
qubits. Ref.~\cite{ref:amp2012} also considered the possibility of generating
more randomness from the joint outcomes of projective measurements performed
on both subsystems. In this case, \cite{ref:amp2012} found that the maximum
of two uniformly random bits could be generated, but only confirmed that this
was attainable using a maximally entangled state
$\ket{\phi_{+}} = \bigro{\ket{00} + \ket{11}}/\sqrt{2}$ or one could get
arbitrarily close to it using a very weakly entangled state of the form
$\ket{\psi_{\theta}} = \cos(\theta/2) \ket{00} + \sin(\theta/2) \ket{11}$ in
the limit $\theta \to 0$ where it becomes separable. Between these two
extremes, determining the amount of randomness that can be generated remains
an open problem.

\begin{table}[t!]
  \centering
  \begin{tabular}{|cc|P{2.5cm}|P{2.5cm}|}
    \cline{3-4}
    \multicolumn{2}{c|}{} & $\ket{\phi_{+}}$ & $\ket{\psi_{\theta}}$ \\
    \hline
    \multirow{2}{*}{Local} & \tiny{PROJ} &
    1 bit \cite{ref:pam2010} &
    1 bit \cite{ref:amp2012} \\
    & \tiny{POVM} & 2 bits \cite{ref:ap2016} &
    2 bits \cellcolor{green!40} \\ \hline
    \multirow{2}{*}{Global} & \tiny{PROJ} &
    2 bits \cite{ref:amp2012} &
    2 bits \cellcolor{green!40} \\
    & \tiny{POVM} & [3.58, 3.96] bits \cellcolor{green!40} &
    [3.58, 3.96] bits \cellcolor{green!40} \\ \hline
  \end{tabular}
  \caption{Maximum amount of randomness (quantified by the min-entropy)
    extractable from one (local) or jointly from two (global) projective
    (PROJ) or nonprojective (POVM) measurements from the maximally
    ($\ket{\phi_{+}}$) and any partially ($\ket{\psi_{\theta}}$) entangled
    two-qubit pure state. Square brackets indicate a range (rounded outward)
    within which the optimal amount of randomness is known to lie. The
    results proved in this work are highlighted. \label{tab:context}}
\end{table}

As well as projective measurements, it is also possible to perform
nonprojective measurements on quantum systems. Nonprojective measurements can
potentially generate more randomness as they can have more outcomes than the
dimension of the quantum system they act on. Extremal qubit measurements in
particular may have up to four outcomes \cite{ref:dalpp2005}. In a bipartite
Bell-type experiment this means that potentially up to two bits of randomness
could be generated locally or up to four bits globally using nonprojective
measurements. The first limit is known to be attainable:
Ref.~\cite{ref:ap2016} describes a way in which two bits of randomness can be
generated locally using a single measurement on one side. But it is currently
an open question whether the second limit of four bits is attainable
globally. The same work, \cite{ref:ap2016}, only confirmed numerically that
at least around 2.8997 bits of randomness can be generated this way. Both
results were established only for the maximally entangled state.

In this work, we solve the question of how much randomness can be generated
using projective measurements from a generic pure entangled state of two
qubits and show that the upper limit of two bits is always attainable
regardless of how strong or weak the entanglement is. We also show that,
alternatively, two bits of randomness can be extracted locally from any such
state using a nonprojective measurement. It turns out however, as we will
detail below, that nonprojective measurements can only be reconstructed to a
limited degree from the correlations observed in a Bell experiment and this
limits the amount of randomness that can be generated globally. We rule out
that any scheme can generate more than about 3.9527 bits of randomness in
this way, proving that the potential upper limit of four bits is not
attainable. We nevertheless show that at least around 3.5850 bits of
randomness can be generated globally with suitable nonprojective measurements
from any partially entangled state.

\rsection{The Bell test} To introduce the problem, we begin by considering
the form of an arbitrary partially entangled state of two qubits. Such a
state can always be expressed in its Schmidt decomposition as
\begin{equation}
  \label{eq:psi_theta}
  \ket{\psi_{\theta}}
  = \cos \bigro{\tfrac{\theta}{2}} \ket{00}
  + \sin \bigro{\tfrac{\theta}{2}} \ket{11}
\end{equation}
for an angle $\theta$ that, without loss of generality, we can and hereafter
will take to be in the range $0 < \theta \leq \tfrac{\pi}{2}$. The same state
is equivalently represented by its density operator
$\psi_{\theta} = \proj{\psi_{\theta}}$, which we can express as
\begin{IEEEeqnarray}{rl}
  \label{eq:psi_theta_density}
  \psi_{\theta} = \frac{1}{4} \Bigsq{& \id \otimes \id + \cos(\theta)
    \bigro{\id \otimes \sz + \sz \otimes \id}
    \nonumber \\
    &+\> \sin(\theta) \bigro{\sx \otimes \sx - \sy \otimes \sy}
    + \sz \otimes \sz}
\end{IEEEeqnarray}
in terms of the identity and Pauli operators $\id$, $\sx$, $\sy$, and $\sz$
acting on each subsystem. We can see that Alice and Bob will have to perform
measurements in the $\sx\text{-}\sy$ plane, for example $A = \sx$ and
$B = \sy$, in order to extract two uniformly random bits from this state,
since this is the only way to have
$\avg{A} = \avg{A \otimes B} = \avg{B} = 0$. We would, however, intuitively
expect the maximum violation of a Bell inequality on $\psi_{\theta}$ to be
attained with measurements having a component in the $\sz$ direction, since
the correlation terms involving $\sz$ in \eqref{eq:psi_theta_density} are
larger in magnitude than the analogous terms involving $\sx$ and $\sy$. As
such, we anticipate that we will need a Bell experiment engineered to exploit
the entire Bloch sphere.

To this end, we propose the following Bell test in which Alice and Bob
perform $\pm 1$-valued measurements $A_{x}$, $x = 1,2,3$ and $B_{y}$,
$y = 1,\dotsc,6$, in each round. They use the statistics to estimate the
values of three Bell expressions. The first two,
\begin{IEEEeqnarray}{rCl}
  \label{eq:Ibeta_def}
  I_{\beta} &=& \avg{\beta A_{1} + A_{1} (B_{1} + B_{2})
    + A_{2} (B_{1} - B_{2})} \,, \IEEEeqnarraynumspace \\
  \label{eq:Jbeta_def}
  J_{\beta} &=& \avg{\beta A_{1} + A_{1} (B_{3} + B_{4})
    + A_{3} (B_{3} - B_{4})} \,,
\end{IEEEeqnarray}
are modified CHSH expressions of the kind introduced in \cite{ref:amp2012},
while the third,
\begin{equation}
  \label{eq:chsh_def}
  S = \avg{A_{2} (B_{5} + B_{6}) + A_{3} (B_{5} - B_{6})} \,,
\end{equation}
is an ordinary CHSH \cite{ref:ch1969,ref:c1980} expression. We choose
\begin{equation}
  \label{eq:beta_eq_ftheta}
  \beta = \frac{2 \cos(\theta)}{\sqrt{1 + \sin(\theta)^{2}}}
\end{equation}
for the value of the parameter $\beta$ in the definitions of $I_{\beta}$ and
$J_{\beta}$, depending on the angle $\theta$ that identifies the intended
state $\ket{\psi_{\theta}}$. Alice and Bob should in particular check that
these Bell expressions attain the values
\begin{IEEEeqnarray}{rCl}
  \label{eq:Ibeta}
  I_{\beta} &=& 2 \sqrt{2} \sqrt{1 + \beta^{2} / 4} \,, \\
  \label{eq:Jbeta}
  J_{\beta} &=& 2 \sqrt{2} \sqrt{1 + \beta^{2} / 4} \,, \\
  \label{eq:chsh}
  S &=& 2 \sqrt{2} \sin(\theta) \,.
\end{IEEEeqnarray}

The Bell expectation values \eqref{eq:Ibeta}, \eqref{eq:Jbeta}, and
\eqref{eq:chsh} can be attained by measuring
\begin{IEEEeqnarray}{c+c+c}
  A_{1} = \sz \,, & A_{2} = \sx \,, & A_{3} = \pm \sy
\end{IEEEeqnarray}
on Alice's side and performing suitable measurements on Bob's side on
$\ket{\psi_{\theta}}$ \cite{ref:amp2012}. Crucially for the intended
application to randomness generation this is, up to trivial modifications
such as local changes of bases and extensions to higher dimension,
essentially the only way to attain these expectation values. More precisely,
in appendix~\ref{sec:XYZ_selftest} we establish the following, which holds
regardless of the Hilbert-space dimension.

\begin{lemma}
  \label{lem:selftest}
  The conditions $I_{\beta} = J_{\beta} = 2\sqrt{2} \sqrt{1 + \beta^{2}/4}$
  and $S = 2 \sqrt{2} \sin(\theta)$ identify an extremal point in the quantum
  set and if they are attained there is a choice of local bases in which:
  \begin{enumerate}[label=\textit{\roman*})]
    \item the underlying state has the form
    \begin{equation}
      \label{eq:selftest_state}
      \rho = \psi_{\theta} \otimes \sigma_{\rA'\rB'} \,,
    \end{equation}
    where $\psi_{\theta}$ is the partially entangled state
    \eqref{eq:psi_theta_density} and $\sigma_{\rA'\rB'}$ is an ancillary
    state which may be of any dimension;
    \item \label{enum:alicemeas}
    Alice's measurements act on the state according to
    \begin{IEEEeqnarray}{rCl}
      A_{1} &=& \sz \otimes \id_{\rA'} \,, \\
      A_{2} &=& \sx \otimes \id_{\rA'} \,, \\
      A_{3} &=& \sy \otimes A'
    \end{IEEEeqnarray}
    where $A'$ is a $\pm 1$-valued Hermitian operator;
    \item \label{enum:bobmeas}
    Bob's measurements act on the state according to
    \begin{IEEEeqnarray}{rCcCl}
      \label{eq:selftestbobZ}
      \frac{B_{1} + B_{2}}{\sqrt{2 \lambda_{+}}}
      &=& \frac{B_{3} + B_{4}}{\sqrt{2 \lambda_{+}}}
      &=& \sz \otimes \id_{\rB'} \,, \\
      \label{eq:selftestbobX}
      \frac{B_{1} - B_{2}}{\sqrt{2 \lambda_{-}}}
      &=& \frac{B_{5} + B_{6}}{\sqrt{2}}
      &=& \sx \otimes \id_{\rB'} \,, \\
      \label{eq:selftestbobY}
      -\frac{B_{3} - B_{4}}{\sqrt{2 \lambda_{-}}}
      &=& -\frac{B_{5} - B_{6}}{\sqrt{2}}
      &=& \sy \otimes B' \,,
    \end{IEEEeqnarray}
    where $\lambda_{\pm} = 1 \pm \beta^{2}/4$ and $B'$ is a $\pm 1$-valued
    Hermitian operator;
    \item \label{enum:sigma_structure} the ancillary state
    $\sigma_{\rA'\rB'}$ in \eqref{eq:selftest_state} and operators $A'$ and
    $B'$ are related in such a way that
    \begin{equation}
      \label{eq:sigma_structure}
      \avg{A' \otimes B'}_{\sigma_{\rA'\rB'}} = 1 \,.
    \end{equation}
  \end{enumerate}
\end{lemma}

The operators $A'$ and $B'$ appearing in lemma~\ref{lem:selftest} are
inevitable and reflect the fact that we cannot distinguish a set of qubit
measurements from their complex conjugates on both sides
\cite{ref:mm2011}. We should also remark that, strictly speaking,
\ref{enum:alicemeas} and \ref{enum:bobmeas} give the form of Alice's and
Bob's measurements only on the supports of the respective marginals
$\rho_{\rA} = \Tr_{\rB}[\rho]$ and $\rho_{\rB} = \Tr_{\rA}[\rho]$ of the
underlying state. This is not a problem for us since any action the
measurements may have on part of the Hilbert space not containing the state
cannot have any impact on the correlations. In the following we will assume,
for simplicity, that the marginals are of full rank.

\rsection{Randomness with projective measurements} Lemma~\ref{lem:selftest}
makes it straightforward to show that we can device-independently extract up
to two bits of randomness using projective measurements. To do this, we
simply add a seventh measurement, $B_{7}$, to the Bell test on Bob's side and
check that its correlation with $A_{2}$ is
\begin{equation}
  \label{eq:global_rand_condition}
  \avg{A_{2} B_{7}} = \sin(\theta) \,.
\end{equation}
Using $A_{2} = \sx \otimes \id_{\rA'}$ and
$\rho = \psi_{\theta} \otimes \sigma_{\rA'\rB'}$ from
lemma~\ref{lem:selftest} and tracing out everything on Alice's side, we can
rewrite the correlation on the left as
\begin{equation}
  \label{eq:corrB7_condA2}
  \avg{A_{2} B_{7}}
  = \sin(\theta) \Tr \bigsq{
    B_{7} \, \tfrac{1}{2} \sx \otimes \sigma_{\rB'}} \,.
\end{equation}
The operator $\tfrac{1}{2} \sx \otimes \sigma_{\rB'}$ has a trace norm of 1
and, since we are assuming $\sigma_{\rB'}$ is of full rank, the only way for
the right sides of \eqref{eq:global_rand_condition} and
\eqref{eq:corrB7_condA2} to be equal is with
\begin{equation}
  B_{7} = \sx \otimes \id_{\rB'} \,.
\end{equation}
With this information we can prove that the results of measuring $A_{3}$ and
$B_{7}$ are maximally random. The probabilities of the four possible outcomes
are
\begin{equation}
  \label{eq:projective_twobits}
  P(ab|37) = \frac{1}{4} \bavg{(\id + a A_{3}) \otimes (\id + b B_{7})} \,,
\end{equation}
$a, b \in \{\pm 1\}$. Direct calculation with $A_{3} = \sy \otimes A'$ and
$B_{7} = \sx \otimes \id$ gives
\begin{equation}
  P(ab|37) = \frac{1}{4} \,.
\end{equation}

Importantly, the fact that we can derive $P(ab|37) = 1/4$ from
$I_{\beta} = J_{\beta} = 2 \sqrt{2} \sqrt{1 + \beta^{2}/4}$,
$S = 2 \sqrt{2} \sin(\theta)$, and $\avg{A_{2} B_{7}} = \sin(\theta)$ shows
that these conditions together are extremal, i.e., they cannot be attained by
averaging quantum strategies that give different values of these
quantities. This rules out the possibility of a more detailed underlying
explanation of the correlations that might allow better predictions to be
made about $A_{3}$ and $B_{7}$.

\rsection{Tomographic reconstruction of POVMs} POVMs performed on qubit
systems can have more than two outcomes and can potentially be used to
generate more randomness than projective measurements. The nature of the
device-independent scenario means we will only be interested in POVMs that
are extremal, i.e., that cannot be expressed as convex combinations of other
POVMs. The extremal qubit POVMs were classified in \cite{ref:dalpp2005} and
the only nontrivial ones consist of at most four rank-one elements
$\alpha_{a} = \proj{\alpha_{a}}$ that are linearly independent.

We can certify the randomness of \emph{some} POVMs device-independently by
using a form of tomography to partially reconstruct them. To see how this
works note first that, in the device-dependent setting, we can reconstruct
any extremal \emph{qubit} POVM $\{\alpha_{a}\}$ on (for example) Alice's side
from the correlations it produces with the Pauli operators on Bob's
side. That is, it turns out that the expectation values
$\avg{\alpha_{a} \otimes \sigma_{\nu}}_{\psi_{\theta}}$, for
$\sigma_{\nu} = \bigro{\id, \sx, \sy, \sz}$ on Bob's side, contain sufficient
information to uniquely identify $\{\alpha_{a}\}$ on Alice's side provided
that the underlying state $\ket{\psi_{\theta}}$ is known.

In the device-independent scenario, we do not know that the quantum system we
are manipulating is limited to a pair of qubits. However, according to
lemma~\ref{lem:selftest} we can verify that Alice is performing Pauli-type
measurements up to complex conjugation, and the linear combinations of Bob's
measurements in \eqref{eq:selftestbobZ}--\eqref{eq:selftestbobY} effectively
give us such operators on Bob's side. With this, we can check that a POVM
$\{R_{a}\}$ on (for example, again) Alice's side produces correlations
consistent with an extremal qubit one, i.e., that
\begin{equation}
  \label{eq:expRaBnu}
  \avg{R_{a} \otimes B_{\nu}}_{\psi_{\theta} \otimes \sigma}
  = \avg{\alpha_{a} \otimes \sigma_{\nu}}_{\psi_{\theta}} \,,
\end{equation}
with
$B_{\nu} = \bigro{\id \otimes \id, \sx \otimes \id, \sy \otimes B', \sz
  \otimes \id}$, where $\{\alpha_{a}\}$ is some ideal reference qubit
POVM. In appendix~\ref{sec:povm_selftest}, we prove that this allows us to
infer the following on the form of $\{R_{a}\}$.
  
\begin{lemma}
  \label{lem:tomography}
  If the correlations obtained from a POVM $\{R_{a}\}$ match those obtainable
  from an extremal reference qubit POVM $\{\alpha_{a}\}$ according to
  \eqref{eq:expRaBnu}, then the elements $R_{a}$ must be of the form
  \begin{IEEEeqnarray}{rCl}
    \label{eq:povm4_tomography}
    R_{a} &=& \alpha_{a} \otimes A'_{+} + \alpha^{*}_{a} \otimes A'_{-}
    \nonumber \\
    &&+\> \trans{\alpha_{a}}{\alpha^{*}_{a}} \otimes K'_{a}
    + \trans{\alpha^{*}_{a}}{\alpha_{a}} \otimes {K'_{a}}^{\dagger} \,,
  \end{IEEEeqnarray}
  where $\ket{\alpha^{*}_{a}}$ is the complex conjugate of
  $\ket{\alpha_{a}}$, $A'_{\pm} = (\id \pm A')/2$ are projectors on the
  positive and negative parts of $A'$ from lemma~\ref{lem:selftest}, and the
  $K'_{a}$ satisfy the operator inequalities
  \begin{IEEEeqnarray}{c+c}
    \label{eq:Ka_opinequalities}
    K'_{a} {K'_{a}}^{\dagger} \leq A'_{+} \,, &
    {K'_{a}}^{\dagger} K'_{a} \leq A'_{-}
  \end{IEEEeqnarray}
  and the condition
  \begin{equation}
    \label{eq:Ka_sumtozero}
    \sum_{a} \trans{\alpha_{a}}{\alpha^{*}_{a}} \otimes K'_{a} = 0 \,.
  \end{equation}
  Furthermore, if $\{R_{a}\}$ has three outcomes or less then $K'_{a} = 0$
  and \eqref{eq:povm4_tomography} simplifies to
  \begin{equation}
    R_{a} = \alpha_{a} \otimes A'_{+} + \alpha^{*}_{a} \otimes A'_{-} \,.
  \end{equation}
\end{lemma}

In other words, our Bell test allows us to reconstruct, up to complex
conjugation, extremal POVMs with two or three outcomes but we can only
partially constrain the form of POVMs with four outcomes. As we elaborate on
in the appendix, the place where the number of outcomes makes a difference is
in the condition \eqref{eq:Ka_sumtozero}: when there are three outcomes or
less, the off-diagonal $\trans{\alpha_{a}}{\alpha^{*}_{a}}$ terms in
\eqref{eq:povm4_tomography} are always linearly independent and thus
\eqref{eq:Ka_sumtozero} can only be satisfied with $K'_{a} = 0$. On the other
hand, a simple calculation shows that $\bra{\alpha^{*}} \sy \ket{\alpha} = 0$
for any qubit state vector; this means that the
$\trans{\alpha_{a}}{\alpha^{*}_{a}}$s are restricted to the three-dimensional
space of operators spanned by $\{\id, \sx, \sz\}$ and they can never be
linearly independent if there are four of them. In that case it is always
possible to satisfy \eqref{eq:Ka_sumtozero} with nonzero $K'_{a}$s.

\rsection{Randomness with POVMs} As we stated earlier, the maximum amount of
randomness that could potentially be generated if both parties use extremal
POVMs is limited to four bits. It is indeed possible to find extremal qubit
POVMs that can generate arbitrarily close to this amount of randomness from
any partially entangled state $\ket{\psi_{\theta}}$. Unfortunately, the fact
that we cannot fully self-test POVMs means that this bound is not attainable
in the device-independent setting. To see this, let us suppose that Alice and
Bob unknowingly try to generate their random results using four-outcome POVMs
$\{R_{a}\}$ and $\{S_{b}\}$ which are related to some ideal extremal qubit
POVMs $\{\alpha_{a}\}$ and $\{\beta_{b}\}$ by
\begin{IEEEeqnarray}{rCl}
  R_{a} &=& \alpha_{a} \otimes \proj{+}_{\rA'}
  + \alpha^{*}_{a} \otimes \proj{-}_{\rA'} \nonumber \\
  &&+\> \lambda_{a} \, \trans{\alpha_{a}}{\alpha^{*}_{a}}
  \otimes \trans{+}{-}_{\rA'} \nonumber \\
  &&+\> \lambda^{*}_{a} \, \trans{\alpha^{*}_{a}}{\alpha_{a}}
  \otimes \trans{-}{+}_{\rA'} \,, \\
  S_{b} &=& \beta_{b} \otimes \proj{+}_{\rB'}
  + \beta^{*}_{b} \otimes \proj{-}_{\rB'} \nonumber \\
  &&+\> \mu_{b} \, \trans{\beta_{b}}{\beta^{*}_{b}}
  \otimes \trans{+}{-}_{\rB'} \nonumber \\
  &&+\> \mu^{*}_{b} \trans{\beta^{*}_{b}}{\beta_{b}}
  \otimes \trans{-}{+}_{\rB'} \,,
\end{IEEEeqnarray}
where $\lambda_{a}$ and $\mu_{b}$ are some complex coefficients with
magnitudes less than 1, while an eavesdropper at each round randomly and
equiprobably chooses and prepares one of two states
$\psi_{\theta} \otimes \chi'_{+}$ or $\psi_{\theta} \otimes \chi'_{-}$ with
different ancillary parts
$\ket{\chi'_{\pm}} = \bigro{\ket{++} \pm \ket{--}} / \sqrt{2}$. Using that
$\braket{\alpha^{*} \beta^{*}}{\psi_{\theta}} = \braket{\alpha
  \beta}{\psi_{\theta}}^{*}$, we can work out that the joint probability of
Alice's and Bob's outcomes, conditioned on either ancillary state being
chosen by Eve, is
\begin{IEEEeqnarray}{l}
  \label{eq:povm_jointprobs}
  \avg{R_{a} \otimes S_{b}}_{\psi_{\theta} \otimes \chi'_{\pm}} \nonumber \\
  \quad = \babs{\braket{\alpha_{a} \beta_{b}}{\psi_{\theta}}}^{2}
  \pm \re \bigsq{\lambda^{*}_{a} \mu^{*}_{b}
    \braket{\alpha_{a} \beta_{b}}{\psi_{\theta}}^{2}} \,.
\end{IEEEeqnarray}
These probabilities average out to the ideal joint probabilities
$\babs{\braket{\alpha_{a} \beta_{b}}{\psi_{\theta}}}^{2}$ that would be
obtained from the reference qubit POVMs on $\ket{\psi_{\theta}}$; hence,
Alice and Bob have no way to detect, device independently, that they are
measuring $\{R_{a}\}$ and $\{S_{b}\}$ rather than $\{\alpha_{a}\}$ and
$\{\beta_{b}\}$. Eve, however, knowing which ancilla state she chose, also
knows which of the two joint distributions in \eqref{eq:povm_jointprobs} was
actually prepared in each round and can use this to make a more informed
guess about what the outcome will be.

Let us see how this could help Eve in the worst case. As we pointed out
above, the off-diagonal terms $\trans{\alpha_{a}}{\alpha^{*}_{a}}$ and
$\trans{\beta_{b}}{\beta^{*}_{b}}$ are never linearly independent and, thus,
the coefficients $\lambda_{a}$ and $\mu_{b}$ can be chosen nonzero. We are
free to scale them such that the largest coefficient on each side is of
magnitude one. By also exploiting the freedom to choose their phases we can
arrange that, for at least one pair $(a, b)$ of outputs,
$\re\bigsq{\lambda^{*}_{a} \mu^{*}_{b} \braket{\alpha_{a}
    \beta_{b}}{\psi_{\theta}}^{2}} = \babs{\braket{\alpha_{a}
    \beta_{b}}{\psi_{\theta}}}^{2}$. In other words, we are certain we can
arrange for at least one of the probabilities
$\avg{R_{a} \otimes S_{b}}_{\psi_{\theta} \otimes \chi'_{-}}$ to be
zero. This means that the probability of the most likely joint outcome,
conditioned on Eve choosing $\ket{\chi'_{-}}$, cannot be lower than
$1/15$. It follows that the randomness that can be certified
device-independently for the entire protocol can never be higher than
\begin{equation}
  -\log_{2} \biggsq{\frac{1}{2} \biggro{\frac{1}{15} + \frac{1}{16}}}
  \approx 3.9527 \text{ bits}
\end{equation}
regardless of the state and POVMs that Alice and Bob try to use.

On a more positive note, the above-described complication does not manifest
if only one of the parties uses a measurement with four outcomes and, in that
case, the amount of randomness that can be generated device-independently is
the same as the amount of randomness that can be generated using extremal
qubit POVMs performed on $\ket{\psi_{\theta}}$. This means it is potentially
possible to generate up to two bits of randomness locally, or alternatively
potentially up to
\begin{equation}
  -\log_{2}(1/12) \approx 3.5850 \text{ bits}
\end{equation}
of randomness globally using a four-outcome POVM on one side and a
three-outcome POVM on the other. We give explicit constructions of POVMs that
yield these amounts of randomness (or arbitrarily close) in
appendix~\ref{sec:extremal_povms}.

\rsection{Conclusion} Our work reinforces the observation that the amount of
randomness obtainable from a quantum system does not in general increase with
the degree of entanglement. In two versions of the problem, we have confirmed
that an upper limit of two bits of randomness is always obtainable from any
partially entangled pure state of two qubits. In the global case using POVMs,
although we do not know the optimal amount of extractable randomness we have
significantly narrowed the range to between about 3.58 and 3.96 bits for any
state. The nontrivial latter limit establishes that the upper bound of four
bits is not attainable in this case.

Our results rely on the fact that we can reconstruct the underlying quantum
state and measurements in our Bell test sufficiently well to conclude that
the outcomes are genuinely random. This adds to a growing literature showing
that we can often infer substantial information about the quantum resources
available from a Bell test
\cite{ref:kt1985,ref:bmr1992,ref:sg2001,ref:my2004,ref:mkys2012}. Previous
work has notably shown that the partially entangled state
\cite{ref:yn2013,ref:bp2015} or measurements spanning the entire Bloch sphere
(up to complex conjugation) \cite{ref:ap2016,ref:k2017,ref:ab2017} can be
self-tested, although before now not together in the same test.

Our work also led us to investigate whether it is possible to self-test
nonprojective measurements in quantum physics and we found that qubit POVMs
with four outcomes can only be self-tested to a limited extent. The ambiguity
with respect to complex conjugation can thus, as we found here, make a
significant difference in the device-independent setting. It will be
interesting to further explore this problem, both for qubit systems and in
higher dimension. In particular, closing the gap on optimal randomness
generation with POVMs is likely to require developing a better understanding
of the general form that we found POVMs may take in
lemma~\ref{lem:tomography}.

\rsection{Note added} While completing this work, we learned that the authors
of~\cite{ref:rpIP} had independently found using a similar approach that two
bits of randomness can be generated globally using projective measurements
from the partially entangled state.

\rsection{Acknowledgements} This work was supported by the Spanish MINECO
(Severo Ochoa grant SEV-2015-0522), the Generalitat de Catalunya (CERCA
Program and SGR 1381), the Fundaci\'o{} Privada Cellex and Mir-Puig, the AXA
Chair in Quantum Information Science, the ERC AdG CERQUTE and CoG QITBOX, and
the EU Quantum Flagship project QRANGE. J.~K.~acknowledges support from the
National Science Centre, Poland (grant no.~2016/23/P/ST2/02122). This project
is carried out under POLONEZ programme which has received funding from the
European Union's Horizon 2020 research and innovation programme under the
Marie Sk{\l}odowska-Curie grant agreement no.~665778. B.~B.~acknowledges
support from the Secretaria d'Universitats i Recerca del Departament
d'Economia i Coneixement de la Generalitat de Catalunya and the European
Social Fund (FEDER). R.~A.~acknowledges support from the Foundation for
Polish Science through the First Team project (First TEAM/2017-4/31)
cofinanced by the European Union under the European Regional Development
Fund.

\bibliography{references}

\clearpage

\appendix

\section{Self-test of state and measurements}
\label{sec:XYZ_selftest}

The core behind our randomness results in the main text is a self-test,
lemma~\ref{lem:selftest}, that we prove here. To recapitulate, we let
\begin{IEEEeqnarray}{rCl}
  \label{eq:Ibeta_def_appx}
  I_{\beta} &=& \avg{\beta A_{1} + A_{1} (B_{1} + B_{2})
    + A_{2} (B_{1} - B_{2})} \,, \IEEEeqnarraynumspace \\
  \label{eq:Jbeta_def_appx}
  J_{\beta} &=& \avg{\beta A_{1} + A_{1} (B_{3} + B_{4})
    + A_{3} (B_{3} - B_{4})} \,, \\
  \label{eq:chsh_def_app}
  S &=& \avg{A_{2} (B_{5} + B_{6}) + A_{3} (B_{5} - B_{6})}
\end{IEEEeqnarray}
denote the expectation values of three Bell operators involving three
$\pm 1$-valued observables $A_{x}$ on Alice's side and six $\pm 1$-valued
observables $B_{y}$ on Bob's side acting on a shared underlying quantum
system. We suppose, furthermore, that the expectation values
\begin{IEEEeqnarray}{rCl}
  \label{eq:Ibeta_appx}
  I_{\beta} &=& 2 \sqrt{2} \sqrt{1 + \beta^{2} / 4} \,, \\
  \label{eq:Jbeta_appx}
  J_{\beta} &=& 2 \sqrt{2} \sqrt{1 + \beta^{2} / 4} \,, \\
  \label{eq:chsh_appx}
  S &=& 2 \sqrt{2} \sin(\theta)
\end{IEEEeqnarray}
are attained, where $\theta$ is related to the parameter $\beta$ by
\begin{equation}
  \label{eq:beta_eq_ftheta_appx}
  \beta = \frac{2 \cos(\theta)}{\sqrt{1 + \sin(\theta)^{2}}} \,.
\end{equation}
According to lemma~\ref{lem:selftest}, this is only possible if there is a
choice of bases in which the quantum state is of the form
\begin{equation}
  \label{eq:selftest_state_appx}
  \rho = \psi_{\theta} \otimes \sigma_{\rA'\rB'} \,,
\end{equation}
where
\begin{IEEEeqnarray}{rl}
  \label{eq:psi_theta_density_appx}
  \psi_{\theta} = \frac{1}{4} \Bigsq{&
    \id \otimes \id
    + \cos(\theta) \bigro{\sz \otimes \id + \id \otimes \sz}
    \nonumber \\
    &+\> \sin(\theta) \bigro{\sx \otimes \sx - \sy \otimes \sy}
    + \sz \otimes \sz} \,,
\end{IEEEeqnarray}
and Alice's and Bob's observables act on the supports of their marginals of
$\rho$ according to
\begin{IEEEeqnarray}{c+c+c}
  A_{1} = \sz \otimes \id \,, &
  A_{2} = \sx \otimes \id \,, &
  A_{3} = \sy \otimes A' \IEEEeqnarraynumspace
\end{IEEEeqnarray}
and
\begin{IEEEeqnarray}{rCcCl}
  \frac{B_{1} + B_{2}}{\sqrt{2 \lambda_{+}}}
  &=& \frac{B_{3} + B_{4}}{\sqrt{2 \lambda_{+}}}
  &=& \sz \otimes \id \,, \\
  \frac{B_{1} - B_{2}}{\sqrt{2 \lambda_{-}}}
  &=& \frac{B_{5} + B_{6}}{\sqrt{2}}
  &=& \sx \otimes \id \,, \\
  -\frac{B_{3} - B_{4}}{\sqrt{2 \lambda_{-}}}
  &=& -\frac{B_{5} - B_{6}}{\sqrt{2}}
  &=& \sy \otimes B' \,,
\end{IEEEeqnarray}
where $A'$ and $B'$ are $\pm 1$-valued Hermitian operators and
$\lambda_{\pm} = 1 \pm \beta^{2}/4$. Furthermore, the ancillary state
$\sigma_{\rA'\rB'}$ in \eqref{eq:selftest_state_appx} and operators $A'$ and
$B'$ are such that
\begin{equation}
  \label{eq:sigma_structure_appx}
  \Tr\bigsq{(A' \otimes B') \sigma_{\rA'\rB'}} = 1 \,.
\end{equation}

We will proceed by deriving constraints on the state and measurements implied
by each of the three Bell expectation values \eqref{eq:Ibeta_appx},
\eqref{eq:Jbeta_appx}, and \eqref{eq:chsh_appx} in turn.

\subsection{$I_{\beta}$ constraint}
\label{sec:I_beta_self_test}

The first condition $I_{\beta} = 2 \sqrt{2} \sqrt{1 + \beta^{2}/4}$ alone
already allows us to self-test the state as well as the measurements $A_{1}$,
$A_{2}$, $B_{1}$, and $B_{2}$. This was essentially proved in the course of
deriving the Tsirelson bound for the more general family of
$I_{\alpha}^{\beta}$ expressions done in \cite{ref:amp2012}, particularly the
steps around Eqs.~(14)--(16). (The result is also closely related to the
self-test based on $I_{\beta}$ in \cite{ref:bp2015}, although the formulation
is slightly different.) Since \cite{ref:amp2012} is not very explicit about
this we will give an alternative and explicit proof here.

$I_{\beta}$ is the expectation value of its corresponding Bell operator,
\begin{equation}
  \mathcal{I}_{\beta} = \beta A_{1} + A_{1} (B_{1} + B_{2})
  + A_{2} (B_{1} - B_{2}) \,.
\end{equation}
Without loss of generality, we take the measurements to be projective, such
that $A\du{x}{2} = B\du{y}{2} = \id$. This allows us to use the Jordan lemma
to decompose the measurements as
\begin{IEEEeqnarray}{rCl}
  A_{x} &=& \sum_{j} A_{x|j} \otimes \proj{j'} \oplus A_{x\bot} \,, \\
  B_{y} &=& \sum_{k} B_{y|k} \otimes \proj{k'} \oplus B_{y\bot} \,,
\end{IEEEeqnarray}
where we absorb any $1 \times 1$ Jordan blocks as well as $2 \times 2$ blocks
where one of the measurements is $\pm \id$ into $A_{x\bot}$ and
$B_{y\bot}$. In other words, $A_{x\bot}$ and $B_{y\bot}$ contain the Jordan
blocks that we already know cannot contribute to a violation of the
$I_{\beta}$ inequality because at least one of the measurements in the blocks
is deterministic. We correspondingly decompose the Bell operator as
\begin{equation}
  \mathcal{I}_{\beta} = \sum_{jk} \mathcal{I}_{\beta|jk} \otimes \proj{j'k'}
  \;\oplus\; \mathcal{I}_{\beta\bot} \,,
\end{equation}
where
\begin{IEEEeqnarray}{rCl}
  \mathcal{I}_{\beta|jk}
  &=& \beta A_{1|j} + A_{1|j} (B_{1|k} + B_{2|k}) \nonumber \\  
  &&+\> A_{2|j} (B_{1|k} - B_{2|k})
\end{IEEEeqnarray}
and $\mathcal{I}_{\beta\bot}$ is the part of $\mathcal{I}_{\beta}$ containing
`$\bot$' blocks on one or the other side or both.

In order to attain the quantum bound, the measurements $A_{x|j}$ and
$B_{y|k}$ in each block must be optimised such that every
$\mathcal{I}_{\beta|jk}$ has an eigenvalue equal to
$2 \sqrt{2} \sqrt{1 + \beta^{2}/4}$ or the underlying state must have zero
presence in the corresponding part of the Hilbert space. From now, let us
redefine $A_{x\bot}$ and $B_{y\bot}$ to also absorb Jordan blocks where the
state has no presence and remove them from further consideration. The
remaining blocks must have measurements optimised to attain the quantum
bound.

Let us concentrate on a particular pair of remaining blocks $(j, k)$ and drop
the indices $jk$ while we do this to lighten the notation. Squaring
$\mathcal{I}_{\beta}$ and using that $A\du{x}{2} = B\du{y}{2} = \id$ gives
\begin{IEEEeqnarray}{rCl}
  \mathcal{I}\du{\beta}{2} &=& (4 + \beta^{2}) \id
  + 2 \beta \id \otimes (B_{1} + B_{2}) \nonumber \\
  &&+\> \beta \acomm{A_{1}}{A_{2}} \otimes (B_{1} - B_{2}) \nonumber \\
  &&-\> \comm{A_{1}}{A_{2}} \otimes \comm{B_{1}}{B_{2}} \,.
\end{IEEEeqnarray}
We are free to choose the bases on both sides in such a way that
\begin{IEEEeqnarray}{rCl}
  A_{1} &=& \sz \,, \\
  A_{2} &=& \cos(\lambda) \sz - \sin(\lambda) \sx \,, \\
  B_{1} + B_{2} &=& 2 \cos \bigro{\tfrac{\mu}{2}} \sz \,, \\
  B_{1} - B_{2} &=& 2 \sin \bigro{\tfrac{\mu}{2}} \sx
\end{IEEEeqnarray}
with all of the $\cos$s and $\sin$s nonnegative. With this choice,
\begin{IEEEeqnarray}{rCl}
  \mathcal{I}\du{\beta}{2}/4 &=& (1 + \beta^{2}/4) \id
  + \beta \cos \bigro{\tfrac{\mu}{2}} \id \otimes \sz \nonumber \\
  &&+\> \beta \cos(\lambda) \sin \bigro{\tfrac{\mu}{2}} \id \otimes \sx
  \nonumber \\
  &&+\> \sin(\lambda) \sin(\mu) \sy \otimes \sy \,.
\end{IEEEeqnarray}
Notice here that only $\id$ and $\sy$, which are simultaneously
diagonalisable, appear on Alice's side. We can use this to identify the
eigenvalues of $\mathcal{I}\du{\beta}{2}/4$. The largest one (with
multiplicity two) is
\begin{IEEEeqnarray}{l}
  \label{eq:Ibeta2_maxeigenvalue}
  1 + \beta^{2}/4 \nonumber \\*
  +\> \sqrt{\beta^{2} \cos \bigro{\tfrac{\mu}{2}}^{2}
  + \beta^{2} \cos(\lambda)^{2} \sin \bigro{\tfrac{\mu}{2}}^{2}
  + \sin(\lambda)^{2} \sin(\mu)^{2}} \,. \nonumber \\*
\end{IEEEeqnarray}
This is maximised with either $\cos(\lambda)^{2} = 1$ or
$\sin(\lambda)^{2} = 1$, depending on whether $\beta^{2} \sin(\mu/2)^{2}$ or
$\sin(\mu)^{2}$ is larger. We can ignore the former since it would mean that
$A_{1}$ and $A_{2}$ commute and we would have already absorbed them into
$A_{x\bot}$, and thus use $\sin(\lambda) = 1$ which fixes $A_{2} = \sx$. The
term under the square root then is
\begin{equation}
  \beta^{2} \cos \bigro{\tfrac{\mu}{2}}^{2}
  + 4 \cos \bigro{\tfrac{\mu}{2}}^{2} \sin \bigro{\tfrac{\mu}{2}}^{2}
\end{equation}
which we can rewrite as
\begin{equation}
  (1 + \beta^{2}/4)^{2}  - \Bigro{2 \cos \bigro{\tfrac{\mu}{2}}^{2}
    - (1 + \beta^{2}/4)}^{2} \,.
\end{equation}
This is maximised with
\begin{equation}
  \cos \bigro{\tfrac{\mu}{2}} = \sqrt{\frac{1 + \beta^{2}/4}{2}} \,,
\end{equation}
which is less than one in the range $\beta < 2$ where $I_{\beta}$ has a
quantum violation, for which the maximum eigenvalues
\eqref{eq:Ibeta2_maxeigenvalue} of $\mathcal{I}\du{\beta}{2}/4$ become
$2 (1 + \beta^{2}/4)$.

Since the other two eigenvalues for the optimal measurements are zero and
$\mathcal{I}_{\beta}$ is traceless we can safely infer that its only two
nonzero eigenvalues are $\pm 2 \sqrt{2} \sqrt{1 + \beta^{2}/4}$ and we can
identify the corresponding eigenstates. Substituting in the
now-known-to-be-optimal qubit measurements, the Bell operator in our Jordan
block is
\begin{IEEEeqnarray}{rCl}
  \label{eq:Ibeta_optmes}
  \mathcal{I}_{\beta} &=& \beta \, \sz \otimes \id
  + \sqrt{2} \sqrt{1 + \beta^{2}/4} \, \sz \otimes \sz \nonumber \\
  &&+\> \sqrt{2} \sqrt{1 - \beta^{2}/4} \, \sx \otimes \sx \,.
\end{IEEEeqnarray}
To identify the eigenstates we express $\mathcal{I}_{\beta}$ in spectral form
as
\begin{equation}
  \label{eq:Ibeta_spectral}
  \mathcal{I}_{\beta}
  = 2 \sqrt{2} \sqrt{1 + \beta^{2}/4} \bigro{\psi_{\theta} - \phi_{\theta}}
\end{equation}
where
\begin{IEEEeqnarray}{rCll}
  \psi_{\theta} &=& \frac{1}{4} \Bigsq{&\id \otimes \id
    + \cos(\theta) \bigro{\sz \otimes \id + \id \otimes \sz} \nonumber \\
    &&&+\> \sin(\theta) \bigro{\sx \otimes \sx - \sy \otimes \sy}
    + \sz \otimes \sz} \,, \\
  \phi_{\theta} &=& \frac{1}{4} \Bigsq{&\id \otimes \id
    - \cos(\theta) \bigro{\sz \otimes \id - \id \otimes \sz} \nonumber \\
    &&&-\> \sin(\theta) \bigro{\sx \otimes \sx + \sy \otimes \sy}
    - \sz \otimes \sz}
\end{IEEEeqnarray}
are rank-one projectors corresponding to pure states
\begin{IEEEeqnarray}{rCl}
  \ket{\psi_{\theta}}
  &=& \cos \bigro{\tfrac{\theta}{2}} \ket{00}
  + \sin \bigro{\tfrac{\theta}{2}} \ket{11} \,, \\
  \ket{\phi_{\theta}}
  &=& \sin \bigro{\tfrac{\theta}{2}} \ket{01}
  - \cos \bigro{\tfrac{\theta}{2}} \ket{10} \,.
\end{IEEEeqnarray}
Precisely, the expressions \eqref{eq:Ibeta_optmes} and
\eqref{eq:Ibeta_spectral} for $\mathcal{I}_{\beta}$ coincide if
\begin{IEEEeqnarray}{c+c}
  \cos(\theta) = \sqrt{\frac{2 \beta^{2}/4}{1 + \beta^{2}/4}} \,, &
  \sin(\theta) = \sqrt{\frac{1 - \beta^{2}/4}{1 + \beta^{2}/4}} \,,
  \IEEEeqnarraynumspace
\end{IEEEeqnarray}
which fixes how $\beta$ and $\mu$ must be related to $\theta$.

Returning back to the full Bell operator, in order to be able to attain its
quantum bound we find that it must be of the form
\begin{equation}
  \mathcal{I}_{\beta}
  = 2 \sqrt{2} \sqrt{1 + \beta^{2}/4} \bigro{\psi_{\theta} - \phi_{\theta}}
  \otimes \id_{\rA'\rB'} \;\oplus\; \mathcal{I}_{\beta\bot} \,,
\end{equation}
and the underlying state must have no support on the
$\mathcal{I}_{\beta\bot}$ part. The problem
\begin{equation}
  \Tr[\mathcal{I}_{\beta} \rho] = 2 \sqrt{2} \sqrt{1 + \beta^{2}/4}
\end{equation}
can thus be solved with, and only with, a state of the form
\begin{equation}
  \rho = \psi_{\theta} \otimes \sigma_{\rA'\rB'} \,.
\end{equation}
To verify this, we express an arbitrary state $\rho$ in spectral form,
\begin{equation}
  \rho = \sum_{k} q_{k} \Psi_{k} \,,
\end{equation}
with $q_{k} > 0$. Then
\begin{equation}
  \Tr[\mathcal{I}_{\beta} \rho]
  = \sum_{k} p_{k} \Tr[\mathcal{I}_{\beta} \Psi_{k}]
  = 2 \sqrt{2} \sqrt{1 + \beta^{2}/4}
\end{equation}
is only possible if
\begin{equation}
  \Tr[\mathcal{I}_{\beta} \Psi_{k}] = 2 \sqrt{2} \sqrt{1 + \beta^{2}/4}
\end{equation}
for all $k$. The left side is more precisely
\begin{IEEEeqnarray}{rCl}
  \Tr[\mathcal{I}_{\beta} \Psi_{k}]
  &=& 2 \sqrt{2} \sqrt{1 + \beta^{2}/4}
  \Tr \bigsq{(\psi_{\theta} - \phi_{\theta}) \otimes \id_{\rA'\rB'} \Psi_{k}}
  \nonumber \\
  &&+\> \Tr[\mathcal{I}_{\beta \bot} \Psi_{k}] \,.
\end{IEEEeqnarray}
From this we can see that, to attain the Tsirelson bound, the second term
$\Tr[\mathcal{I}_{\beta \bot} \Psi_{k}]$ must be zero and we must have
\begin{IEEEeqnarray}{rl}
  \IEEEeqnarraymulticol{2}{l}{
    \Tr \bigsq{(\psi_{\theta} - \phi_{\theta})
      \otimes \id_{\rA'\rB'} \Psi_{k}}} \nonumber \\
  \qquad &= \bra{\psi_{\theta}} \Psi_{2|k} \ket{\psi_{\theta}}
  - \bra{\phi_{\theta}} \Psi_{2|k} \ket{\phi_{\theta}} \nonumber \\
  &= 1 \,,
\end{IEEEeqnarray}
where $\Psi_{2|k} = \Tr_{\rA'\rB'}[\Psi_{k}]$ in the second expression is the
qubit marginal of $\Psi_{k}$. The above problem is only solvable if
$\Psi_{2|k} = \psi_{\theta}$, i.e., if the $\ket{\Psi_{k}}$ are of the form
\begin{equation}
  \ket{\Psi_{k}} = \ket{\psi_{\theta}} \otimes \ket{\chi'_{k}}
\end{equation}
for all $k$. This finally gives
\begin{equation}
  \rho = \psi_{\theta} \otimes \Bigro{\sum_{k} q_{k} \chi'_{k}}
  = \psi_{\theta} \otimes \sigma_{\rA'\rB'} \,.
\end{equation}
Furthermore, the marginal states $\sigma_{\rA'}$ and $\sigma_{\rB'}$ are of
full rank in the Hilbert spaces we implicitly introduced to contain the
Jordan block indices, since we absorbed blocks where the state has no
presence into $\mathcal{I}_{\beta\bot}$.

In the remainder of this proof we will restrict our attention to the Hilbert
space containing the state and take $\rho$ to be such that its marginals
$\rho_{\rA}$ and $\rho_{\rB}$ are of full rank.

\subsection{$J_{\beta}$ constraint}

The second Bell expression,
\begin{equation}
  J_{\beta} = \avg{\beta A_{1} + A_{1} (B_{3} + B_{4})
    + A_{3} (B_{3} - B_{4})} \,,
\end{equation}
is the same as $I_{\beta}$ except with different measurements and the second
condition $J_{\beta} = 2 \sqrt{2} \sqrt{1 + \beta^{2}/4}$ implies analogous
conditions for the measurements, although not necessarily in the same
bases. In particular, the second condition implies that, like $A_{1}$ and
$A_{2}$, $A_{1}$ and $A_{3}$ must anticommute.

Having already identified $\rho$ and fixed $A_{1}$, we can derive
the most general $A_{3}$ that anticommutes with $A_{1}$. In general we may
expand $A_{3}$ as
\begin{equation}
  A_{3} = \id \otimes A'_{\id} + \sx \otimes A'_{\sx}
  + \sy \otimes A'_{\sy} + \sz \otimes A'_{\sz}
\end{equation}
for some Hermitian operators $A'_{\id}$, $A'_{\sx}$, $A'_{\sy}$, and
$A'_{\sz}$. By imposing the conditions $A\du{3}{2} = \id \otimes \id$ and
$\acomm{A_{1}}{A_{3}} = 0$ for $A_{1} = \sz \otimes \id$ on $A_{3}$ we derive
constraints on the $A'$ operators. To satisfy the constraints we find that we
must take $A'_{\id} = A'_{\sz} = 0$ and $A_{3}$ of the form
\begin{equation}
  \label{eq:A3_XY}
  A_{3} = \sx \otimes A'_{\sx} + \sy \otimes A'_{\sy} \,,
\end{equation}
with $A'_{\sx}$ and $A'_{\sy}$ Hermitian and satisfying
\begin{IEEEeqnarray}{rCl}
  {A'_{\sx}}^{2} + {A'_{\sy}}^{2} &=& \id \,, \\
  \comm{A'_{\sx}}{A'_{\sy}} &=& 0 \,.
\end{IEEEeqnarray}
To state the same result differently, $A_{3}$ is of the form
\begin{equation}
  \label{eq:A3_cosXsinY}
  A_{3} = \bigro{\cos(\varphi_{j}) \sx + \sin(\varphi_{j}) \sy}
  \otimes \proj{j'} \,,
\end{equation}
which we obtain by explicitly decomposing the co-diagonal operators
$A'_{\sx}$ and $A'_{\sy}$.

\subsection{$S$ constraint}

As the final step in the proof of our self-test, we will now prove that
satisfying the third condition,
\begin{equation}
  S = \avg{A_{2} (B_{5} + B_{6}) + A_{3} (B_{5} - B_{6})}
  = 2 \sqrt{2} \sin(\theta) \,,
\end{equation}
forces us to set $A_{\sx} = 0$ in \eqref{eq:A3_XY}. We start by writing the
expectation value explicitly as
\begin{equation}
  S = \Tr \bigsq{\bigro{(A_{2} + A_{3}) B_{5} + (A_{2} - A_{3}) B_{5}}
    \, \psi_{\theta} \otimes \sigma_{\rA'\rB'}} \,.
\end{equation}
Using the form \eqref{eq:psi_theta_density_appx} of $\psi_{\theta}$ in terms
of Pauli operators and \eqref{eq:A3_cosXsinY} of $A_{3}$ and doing the trace
on Alice's side first we can rewrite the CHSH expectation value as
\begin{equation}
  S = \sin(\theta) \bigro{\Tr[\rho_{\rB+} B_{5}]
    + \Tr[\rho_{\rB-} B_{6}]} \,,
\end{equation}
where we introduce
\begin{IEEEeqnarray}{rCl}
  \rho_{\rB\pm} &=& \frac{1}{\sin(\theta)}
  \Tr_{\rA} \bigsq{(A_{2} \pm A_{3})
    \, \psi_{\theta} \otimes \sigma_{\rA'\rB'}} \nonumber \\
  &=& \frac{1}{2} \sum_{j} p_{j} \bigro{
    (1 \mp c_{j}) \sx \mp s_{j} \sy}
  \otimes \sigma_{\rB'|j}
\end{IEEEeqnarray}
and, in turn, $c_{j}$ and $s_{j}$ are shorthand for $\cos(\varphi_{j})$ and
$\sin(\varphi_{j})$ and $p_{j}$ and $\sigma_{\rB|j}$ are defined such that
\begin{equation}
  p_{j} \sigma_{\rB'|j}
  = \Tr_{\rA'} \bigsq{(\proj{j'} \otimes \id_{\rB'}) \sigma_{\rA'\rB'}}
\end{equation}
and $\Tr[\sigma_{\rB'|j}] = 1$. Note that this implies $\sum_{j} p_{j} =
1$. We can then upper bound the CHSH expectation value by
\begin{IEEEeqnarray}{rCl}
  \label{eq:S/theta_bound}
  \frac{S}{\sin(\theta)}
  &=& \Tr[\rho_{\rB+} B_{5}] + \Tr[\rho_{\rB-} B_{6}] \nonumber \\
  &\leq& \trnorm{\rho_{\rB+}} + \trnorm{\rho_{\rB-}} \nonumber \\
  &\leq& \sum_{j} p_{j}
  \begin{IEEEeqnarraybox}[][t]{rl}
    \biggro{&
      \frac{1}{2} \btrnorm{(1 - c_{j}) \sx - s_{j} \sy}
      \trnorm{\sigma_{\rB'|j}} \\
      &+\> \frac{1}{2} \btrnorm{(1 + c_{j}) \sx + s_{j} \sy}
      \trnorm{\sigma_{\rB'|j}}}
  \end{IEEEeqnarraybox} \nonumber \\
  &=& \sum_{j} p_{j} \biggro{
    \sqrt{(1 - c_{j})^{2} + {s_{j}}^{2}}
    + \sqrt{(1 + c_{j})^{2} + {s_{j}}^{2}}} \nonumber \\
  &=& \sum_{j} p_{j} \Bigro{\sqrt{2 - 2 c_{j}} + \sqrt{2 + 2 c_{j}}}
  \nonumber \\
  &\leq& \sum_{j} p_{j} \sqrt{2} \sqrt{(2 - 2 c_{j}) + (2 + 2 c_{j})}
  \nonumber \\ 
  &=& 2 \sqrt{2} \,.
\end{IEEEeqnarray}

To determine the optimal $A_{3}$, observe that to attain
\eqref{eq:S/theta_bound} all of the inequalities applied to derive it must
hold with equality. In particular, equality between the third and second last
lines requires
\begin{equation}
  \sqrt{2 + 2 c_{j}} = \sqrt{2 - 2 c_{j}} \,,
\end{equation}
or $c_{j} = 0$ and $s_{j} = \pm 1$. We deduce that
\begin{equation}
  A_{3} = \sy \otimes A'
\end{equation}
where $A' = A'_{+} - A'_{-}$ is the difference of two orthogonal projectors
that sum to the identity. We can then extract $B_{5}$ and $B_{6}$ by
reevaluating CHSH with
\begin{IEEEeqnarray}{rCl}
  A_{2} &=& \sx \otimes (A'_{+} + A'_{-}) \,, \\
  A_{3} &=& \sy \otimes (A'_{+} - A'_{-}) \,.
\end{IEEEeqnarray}
The result can be written
\begin{equation}
  \frac{S}{\sin(\theta)}
  = \tfrac{1}{2} \Tr[\rho_{\rB+} B_{5}]
  + \tfrac{1}{2} \Tr[\rho_{\rB-} B_{6}]
\end{equation}
where
\begin{equation}
  \rho_{\rB'\pm}
  = p_{+} (\sx \mp \sy) \otimes \sigma_{\rB'|+}
  + p_{-} (\sx \pm \sy) \otimes \sigma_{\rB'|-}
\end{equation}
and in turn
\begin{equation}
  p_{\pm} \sigma_{\rB'|\pm} = \Tr_{\rA'}[A'_{\pm} \sigma_{\rA'\rB'}]
\end{equation}
are Bob's partial traces conditioned on projection on the $A'_{\pm}$
subspaces on Alice's side. Satisfying the CHSH constraint requires that
$B_{5}$, $B_{6}$, and $\sigma_{\rB'|\pm}$ be such that
\begin{IEEEeqnarray}{rCcCl}
  \tfrac{1}{2} \Tr[\rho_{\rB+} B_{5}]
  &=& \tfrac{1}{2} \trnorm{\rho_{\rB+}} &=& \sqrt{2} \,, \\
  \tfrac{1}{2} \Tr[\rho_{\rB-} B_{5}]
  &=& \tfrac{1}{2} \trnorm{\rho_{\rB-}} &=& \sqrt{2} \,.
\end{IEEEeqnarray}
This is only possible if $\sigma_{\rB'\pm}$ have orthogonal support and
\begin{IEEEeqnarray}{rCl}
  B_{5} &=& \frac{\sx - \sy}{\sqrt{2}} \otimes B'_{+}
  + \frac{\sx + \sy}{\sqrt{2}} \otimes B'_{-} \,, \\
  B_{6} &=& \frac{\sx + \sy}{\sqrt{2}} \otimes B'_{+}
  + \frac{\sx - \sy}{\sqrt{2}} \otimes B'_{-} \,,
\end{IEEEeqnarray}
where $B'_{\pm}$ are orthogonal projectors on the supports of
$\sigma_{\rB'|\pm}$. The above forms for $B_{5}$ and $B_{6}$ can
alternatively be written
\begin{IEEEeqnarray}{rCl}
  B_{5} &=& \frac{1}{\sqrt{2}} \bigro{
    \sx \otimes \id - \sy \otimes B'} \,, \\
  B_{6} &=& \frac{1}{\sqrt{2}} \bigro{
    \sx \otimes \id + \sy \otimes B'}
\end{IEEEeqnarray}
with $B' = B'_{+} - B'_{-}$, and orthogonality of $\sigma_{\rB'|\pm}$
compactly as the condition
\begin{equation}
  \Tr \bigsq{(A' \otimes B') \sigma_{\rA'\rB'}} = 1 
\end{equation}
on $\sigma_{\rA'\rB'}$.

Similarly, computing the previous Bell expression, $J_{\beta}$, with
$A_{1} = \sz \otimes \id_{\rA'}$, $A_{3} = \sy \otimes A'$, and
$\rho = \psi_{\theta} \otimes \sigma_{\rA'\rB'}$ allows the optimal
measurements $B_{3}$ and $B_{4}$ to be identified as those given at the
beginning of this section.

\section{Self-testing extremal POVMs}
\label{sec:povm_selftest}

The self-test we have described, which among other things allows the Pauli
measurements $\sx$, $\sy$, and $\sz$ to be identified up to complex
conjugation, allows us to perform a form of tomography with nonprojective
measurements. It is stated as lemma~\ref{lem:tomography} in the main text. To
recall the problem: we suppose that Alice performs a POVM $\{R_{a}\}$ and Bob
measures
\begin{equation}
  B_{\nu} = \bigro{\id, \sx \otimes \id, \sy \otimes B', \sz \otimes \id}
\end{equation}
on the state $\psi_{\theta} \otimes \sigma_{\rA'\rB'}$ from our Bell test,
and we check that the correlations obtained match those that could be
obtained from an ideal reference qubit POVM, i.e., that
\begin{equation}
  \label{eq:expRaBnu_apx}
  \avg{R_{a} \otimes B_{\mu}}_{\psi_{\theta} \otimes \sigma}
  = \avg{\alpha_{a} \otimes \sigma_{\nu}}_{\psi_{\theta}} \,,
\end{equation}
where $\{\alpha_{a}\}$ is a given extremal qubit POVM and
$\sigma_{\nu} = (\id, \sx, \sy, \sz)$. We will prove here that this implies
that the POVM elements $R_{a}$ must be of the form asserted in
lemma~\ref{lem:tomography} in the main text, i.e.,
\begin{IEEEeqnarray}{rCl}
  R_{a} &=& \alpha_{a} \otimes A'_{+} + \alpha^{*}_{a} \otimes A'_{-}
  \nonumber \\
  &&+\> \trans{\alpha_{a}}{\alpha^{*}_{a}} \otimes K'_{a}
  + \trans{\alpha^{*}_{a}}{\alpha_{a}} \otimes {K'_{a}}^{\dagger} \,,
\end{IEEEeqnarray}
where $\ket{\alpha^{*}_{a}}$ is the complex conjugate of $\ket{\alpha_{a}}$,
$A'_{\pm} = (\id \pm A')/2$, and the operators $K'_{a}$ are such that
\begin{IEEEeqnarray}{c+c}
  \label{eq:Ka_opinequalities_appx}
  K'_{a} {K'_{a}}^{\dagger} \leq A'_{+} \,, &
  {K'_{a}}^{\dagger} K'_{a} \leq A'_{-}
\end{IEEEeqnarray}
and
\begin{equation}
  \label{eq:Ka_sumtozero_appx}
  \sum_{a} \trans{\alpha_{a}}{\alpha^{*}_{a}} \otimes K'_{a} = 0 \,.
\end{equation}

To begin with, we remark that we can always decompose the reference qubit
POVM $\{\alpha_{a}\}$ in the basis of the identity and Pauli operators as
\begin{equation}
  \label{eq:alpha_decomposition}
  \alpha_{a} = r_{a}^{\mu} \sigma_{\mu}
\end{equation}
for some coefficients $r_{a}^{\mu}$, where we use implicit summation over the
repeated Greek index $\mu$. The correlation on the left side of
\eqref{eq:expRaBnu_apx} can be written as
\begin{equation}
  \label{eq:alpha_sigma_corr}
  \avg{\alpha_{a} \otimes \sigma_{\nu}} = \eta_{\mu\nu} r_{a}^{\mu}
\end{equation}
where $\eta_{\mu\nu} = \avg{\sigma_{\mu} \otimes \sigma_{\nu}}$ make up the
components of a matrix that can be read off the expression
\eqref{eq:psi_theta_density_appx} of $\psi_{\theta}$ in terms of the Pauli
operators. Importantly, for $\theta \neq 0$, the matrix $(\eta_{\mu\nu})$ is
invertible (for example, its determinant is $-\sin(\theta)^{4}$), so that the
correlations $\avg{\alpha_{a} \otimes \sigma_{\nu}}$ uniquely identify the
coefficients $r_{a}^{\nu}$ and the POVM elements $\alpha_{a}$. We can then
combine \eqref{eq:alpha_decomposition} and \eqref{eq:alpha_sigma_corr} to
express the POVM elements directly in terms of the correlations as
\begin{equation}
  \label{eq:alpha_corr_decomp}
  \alpha_{a}
  = \sigma^{\nu} \avg{\alpha_{a} \otimes \sigma_{\nu}}_{\psi_{\theta}} \,,
\end{equation}
where $\sigma^{\nu}$ are operators defined to be related to $\sigma_{\mu}$ by
$\eta_{\mu \nu} \sigma^{\nu} = \sigma_{\mu}$.

By hypothesis, the correlations in \eqref{eq:alpha_corr_decomp} are the same
as those obtained with $R_{a}$ and $B_{\nu}$ according to
\eqref{eq:expRaBnu_apx}; we can thus reexpress $\alpha_{a}$ as
\begin{equation}
  \alpha_{a}
  = \sigma^{\nu} \avg{R_{a} \otimes B_{\nu}}_{
    \psi_{\theta} \otimes \sigma_{\rA'\rB'}} \,.
\end{equation}
Now we note that we can always decompose $R_{a}$ as
\begin{equation}
  R_{a} = \sigma_{\mu} \otimes {R'}_{a}^{\mu}
\end{equation}
and we can write the operators $B_{\nu}$ together as
\begin{equation}
  B_{\nu} = \sigma_{\nu} \otimes B'_{+}
  + \sigma_{\nu}^{*} \otimes B'_{-} \,.
\end{equation}
Using these and developing,
\begin{IEEEeqnarray}{rCl}
  \alpha_{a}
  &=& \sigma^{\nu} \avg{\sigma_{\mu} \otimes \sigma_{\nu}}_{\psi_{\theta}}
  \avg{{R'}_{a}^{\mu} \otimes B'_{+}}_{\sigma_{\rA'\rB'}} \nonumber \\
  &&+\>
  \sigma^{\nu} \avg{\sigma_{\mu} \otimes \sigma_{\nu}^{*}}_{\psi_{\theta}}
  \avg{{R'}_{a}^{\mu} \otimes B'_{-}}_{\sigma_{\rA'\rB'}} \nonumber \\
  &=& \sigma^{\nu} \eta_{\mu\nu}
  \avg{A'_{+} {R'}_{a}^{\mu} A'_{+} \otimes \id_{\rB'}}_{\sigma_{\rA'\rB'}}
  \nonumber \\
  &&+\>
  \sigma^{\nu *} \eta_{\mu\nu}
  \avg{A'_{-} {R'}_{a}^{\mu} A'_{-} \otimes \id_{\rB'}}_{\sigma_{\rA'\rB'}}
  \nonumber \\
  &=& \sigma_{\mu} \Tr \bigsq{{R'}^{\mu}_{a} \, A'_{+} \sigma_{\rA'} A'_{+}}
  \nonumber \\
  &&+\> \sigma_{\mu}^{*} \Tr \bigsq{
    {R'}^{\mu}_{a} \, A'_{-} \sigma_{\rA'} A'_{-}} \,,
\end{IEEEeqnarray}
where we used that $\sigma^{\nu *} = \pm \sigma^{\nu}$ in the same way as
$\sigma_{\nu}^{*} = \pm \sigma_{\nu}$ and the property
\begin{equation}
  A'_{\pm} \otimes B'_{\mp} \, \sigma_{\rA'\rB'} = 0
\end{equation}
to get to the second expression.

Let us now introduce diagonal decompositions
\begin{IEEEeqnarray}{rCl}
  A'_{+} \sigma_{\rA'} A'_{+} &=& \sum_{k} p_{k+} \proj{k'_{+}} \,, \\
  A'_{-} \sigma_{\rA'} A'_{-} &=& \sum_{l} p_{l-} \proj{l'_{-}} \,,
\end{IEEEeqnarray}
with $\sum_{k} p_{k+} + \sum_{l} p_{l-} = 1$, of the two orthogonal operators
$A'_{\pm} \sigma_{\rA'} A'_{\pm}$ appearing in the traces. We then obtain
\begin{equation}
  \label{eq:alpha_extremal_decomp}
  \alpha_{a} = \sum_{k} p_{k+} \, R_{a|k+} + \sum_{l} p_{l-} \, R_{a|l-}^{*}
\end{equation}
where one can verify that
\begin{IEEEeqnarray}{rCl}
  R_{a|k+} &=& \sigma_{\mu} \bra{k'_{+}} {R'}_{a}^{\mu} \ket{k'_{+}} \,, \\
  R_{a|l-} &=& \sigma_{\mu} \bra{l'_{-}} {R'}_{a}^{\mu} \ket{l'_{-}}
\end{IEEEeqnarray}
are POVM elements. Since $\{\alpha_{a}\}$ is by hypothesis extremal,
\eqref{eq:alpha_extremal_decomp} is only possible if
\begin{equation}
  R_{a|k+} = R_{a|l-}^{*} = \alpha_{a} \,.
\end{equation}
Put differently, this means that our uncharacterised POVM $\{R_{a}\}$ must
be such that
\begin{IEEEeqnarray}{rCl}
  \label{eq:Rdiag+constraint}
  \Tr_{\rA'} \bigsq{R_{a} (\id \otimes \proj{k'_{+}})}
  &=& \alpha_{a} \,, \\
  \label{eq:Rdiag-constraint}
  \Tr_{\rA'} \bigsq{R_{a} (\id \otimes \proj{l'_{-}})}
  &=& \alpha_{a}^{*} \,.
\end{IEEEeqnarray}

To further constrain $\{R_{a}\}$, we introduce states
\begin{equation}
  \ket{\varphi'_{\pm}} = \frac{1}{\sqrt{2}}
  \bigro{\ket{j'_{+}} \pm \ket{k'_{+}}} \,,
\end{equation}
with $j < k$, and compute
\begin{IEEEeqnarray}{l}
  \label{eq:Ra_sandwich}
  (\id \otimes \bra{\varphi'_{\pm}}) R_{a} (\id \otimes \ket{\varphi'_{\pm}})
  \nonumber \\
  \quad = \alpha_{a} \pm \Tr_{\rA'} \bigsq{ R_{a} (\id \otimes
    \sx^{+}_{jk})} \,,
\end{IEEEeqnarray}
where
\begin{equation}
  \sx^{+}_{jk} = \frac{1}{2} \Bigro{\trans{j'_{+}}{k'_{+}}
    + \trans{k'_{+}}{j'_{+}}} \,.
\end{equation}
The left side of \eqref{eq:Ra_sandwich} is by construction positive
semidefinite, which implies
\begin{equation}
  \pm \Tr_{\rA'} \bigsq{R_{a} (\id \otimes \sx^{+}_{jk})}
  \leq \alpha_{a} \,.
\end{equation}
Since $\alpha_{a}$ is of rank one, this is only possible if the partial trace
term is itself a multiple of $\alpha_{a}$, i.e., if
\begin{equation}
  \Tr_{\rA'} \bigsq{R_{a} (\id \otimes \sx^{+}_{jk})}
  = \lambda_{a} \alpha_{a} \,.
\end{equation}
Since $\{R_{a}\}$ is a POVM and $\sum_{a} R_{a} = \id$,
\begin{equation}
  \sum_{a} \Tr_{\rA'} \bigsq{R_{a} (\id \otimes \sx^{+}_{jk})}
  = \sum_{a} \lambda_{a} \alpha_{a} = 0 \,.
\end{equation}
Linear independence of the $\alpha_{a}$s means that the second part of this
equality can only be solved with $\lambda_{a} = 0$, from which we obtain the
constraint
\begin{equation}
  \label{eq:RX+constraint}
  \Tr_{\rA'} \bigsq{R_{a} (\id \otimes \sx^{+}_{jk})} = 0 \,.
\end{equation}
Repeating this reasoning starting with states
\begin{equation}
  \ket{\phi'_{\pm}}
  = \frac{1}{\sqrt{2}} \bigro{\ket{l'_{-}} \pm \ket{m'_{-}}} \,,
\end{equation}
and
\begin{IEEEeqnarray}{rCl}
  \ket{\phi'_{\pm}}
  &=& \frac{1}{\sqrt{2}} \bigro{\ket{j'_{+}} \pm i \ket{k'_{+}}} \,, \\
  \ket{\phi'_{\pm}}
  &=& \frac{1}{\sqrt{2}} \bigro{\ket{l'_{-}} \pm i \ket{m'_{-}}} \,,
\end{IEEEeqnarray}
we obtain analogous constraints
\begin{IEEEeqnarray}{rCl}
  \Tr_{\rA'} \bigsq{R_{a} (\id \otimes \sx^{-}_{lm})} &=& 0 \,, \\
  \Tr_{\rA'} \bigsq{R_{a} (\id \otimes \sy^{+}_{jk})} &=& 0 \,, \\
  \Tr_{\rA'} \bigsq{R_{a} (\id \otimes \sy^{-}_{lm})} &=& 0
\end{IEEEeqnarray}
for operators
\begin{IEEEeqnarray}{rCl}
  \label{eq:RX-constraint}
  \sx^{-}_{lm} &=& \frac{1}{2} \Bigro{\trans{l'_{+}}{m'_{+}}
    + \trans{m'_{-}}{l'_{-}}} \,, \\
  \label{eq:RY+constraint}
  \sy^{+}_{jk} &=& \frac{1}{2} \Bigro{-i \trans{j'_{+}}{k'_{+}}
    + i \trans{k'_{+}}{j'_{+}}} \,, \\
  \label{eq:RY-constraint}
  \sy^{-}_{lm} &=& \frac{1}{2} \Bigro{-i \trans{l'_{-}}{m'_{-}}
    + i \trans{m'_{-}}{l'_{-}}} \,.
\end{IEEEeqnarray}

The sets of operators $\{\proj{k'_{+}}, \sx^{+}_{jk}, \sy^{+}_{jk}\}$ and
$\{\proj{l'_{-}}, \sx^{-}_{lm}, \sy^{-}_{lm}\}$ are bases of the spaces of
Hermitian operators acting on the supports of $A'_{+}$ and $A'_{-}$, so the
constraints \eqref{eq:Rdiag+constraint}, \eqref{eq:Rdiag-constraint}, and
\eqref{eq:RX+constraint}--\eqref{eq:RY-constraint} can be written together as
the single operator constraint
\begin{equation}
  \label{eq:Ra_Apmconstraint}
  \id \otimes \$'(R_{a})
  = \alpha_{a} \otimes A'_{+} + \alpha_{a}^{*} \otimes A'_{-} \,,
\end{equation}
where the superoperator $\$'$ acts on operators on $\hilb_{\rA'}$ as
\begin{equation}
  \$'(X') = A'_{+} X' A'_{+} + A'_{-} X' A'_{-} \,.
\end{equation}
The most general operator $R_{a}$ that satisfies \eqref{eq:Ra_Apmconstraint}
is one of the form
\begin{equation}
  R_{a} = \alpha_{a} \otimes A'_{+} + \alpha_{a}^{*} \otimes A'_{-}
  + K_{a} + K_{a}^{\dagger}
\end{equation}
where the off-diagonal operator $K_{a}$ is identified by whatever
\begin{equation}
  K_{a} = (\id \otimes A'_{+}) R_{a} (\id \otimes A'_{-})
\end{equation}
is.

Now we reintroduce that $R_{a}$ is supposed to be a POVM element. The
property $R_{a} \geq 0$ means, by definition, that we must have
$\bra{\phi} R_{a} \ket{\phi} \geq 0$ for any state. Imposing this for certain
choice states allows us to further constrain the form of $R_{a}$. Dropping in
the following the subscript `$a$' for readability, we start with a family of
states
\begin{equation}
  \ket{\phi}
  = \frac{1}{\sqrt{2}} \Bigro{
    \ket{\alpha_{\bot}} \ket{+'}
    + e^{i\varphi} \ket{\alpha^{*}_{\bot}} \ket{-'}} \,,
\end{equation}
where $\varphi$ is some phase, $\ket{\alpha_{\bot}}$ and
$\ket{\alpha^{*}_{\bot}}$ are the unique (up to a phase) qubit states
orthogonal to $\ket{\alpha}$ and $\ket{\alpha^{*}}$, and $\ket{\pm'}$ are
any states in the support of $A'_{\pm}$, such that
$A'_{\pm} \ket{\pm'} = \ket{\pm'}$ and $A'_{\mp} \ket{\pm'} = 0$. For the
minimising phase $\varphi = \varphi_{0}$ we get
\begin{IEEEeqnarray}{rCl}
  \bra{\phi} R \ket{\phi}
  &=& \re\bigsq{
    e^{i\varphi_{0}}
    \bra{\alpha_{\bot}} \bra{+'} K \ket{\alpha^{*}_{\bot}} \ket{-'}}
  \nonumber \\
  &=& - \babs{
    \bra{\alpha_{\bot}} \bra{+'} K \ket{\alpha^{*}_{\bot}} \ket{-'}} \,,
\end{IEEEeqnarray}
from which we conclude that
$\bra{\alpha_{\bot}} \bra{+'} K \ket{\alpha^{*}_{\bot}} \ket{-'} = 0$. Since
we also have $\bra{\alpha_{\bot}} \bra{-'} K = 0$ and
$K \ket{\alpha^{*}_{\bot}} \ket{+'}$, we can generalise this to the operator
equality
\begin{equation}
  (\bra{\alpha_{\bot}} \otimes \id_{\rA'})
  K (\ket{\alpha^{*}_{\bot}} \otimes \id_{\rA'})
  = 0 \,.
\end{equation}
Next, for
\begin{equation}
  \ket{\phi} = \cos \bigro{\tfrac{\theta}{2}} \ket{\alpha} \ket{+'}
  + e^{i\varphi} \sin \bigro{\tfrac{\theta}{2}}
  \ket{\alpha^{*}_{\bot}} \ket{-'}
\end{equation}
and the minimising $(\theta, \varphi) = (\theta_{0}, \varphi_{0})$,
\begin{IEEEeqnarray}{rCl}
  \bra{\phi} R \ket{\phi}
  &=& \frac{1 + \cos(\theta_{0})}{2} \norm{\alpha}^{2}
  \nonumber \\
  &&+\> \sin(\theta_{0}) \re \bigsq{
    e^{i\varphi_{0}}
    \bra{\alpha} \bra{+'} K \ket{\alpha^{*}_{\bot}} \ket{-'}}
  \nonumber \\
  &=& \frac{1}{2} \norm{\alpha}^{2}
  - \frac{1}{2} \sqrt{\norm{\alpha}^{4}
    + 4 \babs{
      \bra{\alpha} \bra{+'} K \ket{\alpha^{*}_{\bot}} \ket{-'}}^{2}} \,,
  \nonumber \\*
\end{IEEEeqnarray}
where $\norm{\alpha} = \Tr[\alpha]$. The second expression is negative unless
$\bra{\alpha} \bra{+'} K \ket{\alpha^{*}_{\bot}} \ket{-'} = 0$; as above,
this lets us conclude
\begin{equation}
  (\bra{\alpha} \otimes \id_{\rA'})
  K (\ket{\alpha^{*}_{\bot}} \otimes \id_{\rA'})
  = 0 \,.
\end{equation}
Similarly, starting with a state
\begin{equation}
  \ket{\phi} = \cos \bigro{\tfrac{\theta}{2}} \ket{\alpha_{\bot}} \ket{+'}
  + e^{i\varphi} \sin \bigro{\tfrac{\theta}{2}} \ket{\alpha^{*}} \ket{-'}
\end{equation}
we obtain
\begin{equation}
  (\bra{\alpha_{\bot}} \otimes \id_{\rA'})
  K (\ket{\alpha^{*}} \otimes \id_{\rA'})
  = 0 \,.
\end{equation}

These three constraints on $K$ mean that the only remaining possibility
is that it is of the form
\begin{equation}
  K = \trans{\alpha}{\alpha^{*}} \otimes K' \,,
\end{equation}
and that $R$ is of the form
\begin{IEEEeqnarray}{rCl}
  \label{eq:Ra_generalform}
  R &=& \alpha \otimes A'_{+} + \alpha^{*} \otimes A'_{-}
  \nonumber \\
  &&+\> \trans{\alpha}{\alpha^{*}} \otimes K'
  + \trans{\alpha^{*}}{\alpha} \otimes {K'}^{\dagger}
\end{IEEEeqnarray}
for some operator $K'$ taking states from the support of $A'_{-}$ to the
support of $A'_{+}$, i.e., satisfying
\begin{equation}
  \label{eq:A+KA-_eq_K}
  A'_{+} K' = K' A'_{-} = K'
\end{equation}
and
\begin{equation}
  \label{eq:A-KA+_eq_K}
  A'_{-} K' = K' A'_{+} = 0 \,.
\end{equation}
We can finally ensure that $R$ is positive semidefinite by requiring that all
of its eigenvalues are nonnegative. We write
\begin{equation}
  \label{eq:Ksvd}
  K' = \sum_{j} \kappa_{j} \trans{j'_{+}}{j'_{-}} \,,
\end{equation}
where the $\kappa_{j} > 0$ are singular values of $K'$ and $\ket{j'_{+}}$ and
$\ket{j'_{-}}$ are states in the supports of $A'_{+}$ and $A'_{-}$. It is
then a straightforward exercise to check that
\begin{equation}
  \ket{\phi} = \ket{\alpha} \ket{j'_{+}} \pm \ket{\alpha^{*}} \ket{j'_{-}}
\end{equation}
are eigenstates of $R$ with the form \eqref{eq:Ra_generalform} above with
eigenvalues
\begin{equation}
  \norm{\alpha} (1 \pm \kappa_{j}) \,,
\end{equation}
from which we extract $\kappa_{j} \leq 1$. This and the conditions
\eqref{eq:A+KA-_eq_K} and \eqref{eq:A-KA+_eq_K} above together are equivalent
to the operator inequalities
\begin{IEEEeqnarray}{c+t+c}
  K' {K'}^{\dagger} \leq A'_{+} &and& {K'}^{\dagger} K' \leq A'_{-} \,.
  \IEEEeqnarraynumspace
\end{IEEEeqnarray}
Conversely, any $R$ that satisfies these conditions is necessarily positive
semidefinite since the only other possible eigenvalues of $R$ are
$\norm{\alpha}$, associated to possible additional eigenstates of $A'_{+}$ or
$A'_{-}$.

Reintroducing the subscript `$a$', the only remaining requirement for
$\{R_{a}\}$ with the form above to be a valid POVM is $\sum_{a} R_{a} = \id$;
this translates directly to the second condition,
\begin{equation}
  \label{eq:Ka_sum_constraint}
  \sum_{a} \trans{\alpha_{a}}{\alpha^{*}_{a}} \otimes K'_{a} = 0 \,,
\end{equation}
stated at the beginning of this section. This is as far as we can go toward
identifying $\{R_{a}\}$ if there are four outcomes. As we pointed out in the
main text, $\Tr \bigsq{\trans{\alpha_{a}}{\alpha^{*}_{a}} \sy} = 0$ and the
operators $\trans{\alpha_{a}}{\alpha^{*}_{a}}$ are thus linear combinations
of $\id$, $\sx$, and $\sz$ and cannot be linearly independent if there are
four of them. In that case it is always possible to satisfy
\eqref{eq:Ka_sum_constraint} with nonzero $K'_{a}$s.

On the other hand, the operators $\trans{\alpha_{a}}{\alpha_{a}^{*}}$ are
always linearly independent if there are no more than three of them provided
that the $\alpha_{a}$ are linearly independent. We can see this from the fact
that one can always change the basis such that all three $\alpha_{a}$ are
real so that $\trans{\alpha_{a}}{\alpha_{a}^{*}} = \alpha_{a}$. To be
precise, let $U$ be a unitary such that, for example,
$U \alpha_{1} U^{\dagger}$ and $U \alpha_{2} U^{\dagger}$ are real. In terms
of the Bloch expressions
\begin{equation}
  \alpha_{a}
  = \frac{1}{2} \norm{\alpha_{a}} \bigro{\id + \vect{n}_{a} \cdot \sv}
\end{equation}
of $\alpha_{a}$, this is any unitary that rotates the normal vector
$\vect{n}_{1} \times \vect{n}_{2}$ onto the $\ry$ axis, which necessarily
rotates $\vect{n}_{1}$ and $\vect{n}_{2}$ onto the $\rx$-$\rz$ plane. Then,
necessarily,
$U \alpha_{3} U^{\dagger} = \id - U \alpha_{1} U^{\dagger} - U \alpha_{2}
U^{\dagger}$ is also real. Applied to the kets, $U \ket{\alpha_{a}}$ are real
up to global phases which are a matter of convention, although we allow for
them explicitly anyway. This means that
\begin{equation}
  e^{-i\varphi_{a}} U \ket{\alpha_{a}}
  = e^{i\varphi_{a}} U^{*} \ket{\alpha_{a}^{*}}
\end{equation}
are real for some phases $\varphi_{a}$. We can use this to relate
$\trans{\alpha_{a}}{\alpha_{a}^{*}}$ to $\alpha_{a}$ by
\begin{equation}
  \label{eq:Uaa*U=aa}
  e^{-2i\varphi_{a}} U \trans{\alpha_{a}}{\alpha_{a}^{*}} U^{T}
  = \alpha_{a} \,.
\end{equation}
This tells us that the $\trans{\alpha_{a}}{\alpha_{a}^{*}}$ must necessarily
be linearly independent if the $\alpha_{a}$ are, since if there were nonzero
parameters $\lambda'_{a}$ such that
$\sum_{a} \lambda'_{a} \trans{\alpha_{a}}{\alpha_{a}^{*}} = 0$ then
\eqref{eq:Uaa*U=aa} would imply that $\sum_{a} \lambda_{a} \alpha_{a} = 0$
with the nonzero parameters $\lambda_{a} = e^{2i\varphi_{a}} \lambda'_{a}$.

\section{Randomness generation with POVMs}
\label{sec:extremal_povms}

Having identified the extent to which we can self-test extremal qubit POVMs
we can return to our original problem of randomness generation. Here we prove
that the parties can generate two bits of randomness locally and arbitrarily
close to $\log_{2}(12) \approx 3.5850$ bits of randomness globally by adding
POVMs to the Bell test. Before confirming that there are indeed POVMs that
yield these amounts of randomness we verify, provided that only one of the
parties uses a four-outcome measurement, that the randomness of the outcomes
is the same as the randomness yielded by reference qubit POVMs.

\subsection{Reduction to qubits}

In an adversarial scenario, Alice and Bob may share an extension
$\rho_{\rABE}$ of their quantum state prepared by an eavesdropper, Eve. From
the Bell test, we can infer that such a state must have the form
\begin{equation}
  \rho_{\rABE} = \psi_{\theta} \otimes \sigma_{\rA'\rB'\rE} \,,
\end{equation}
where $\sigma_{\rA'\rB'\rE}$ is an extension of the ancillary state
$\sigma_{\rA'\rB'}$ from the self-test. By performing a measurement
$\{\Pi_{e}\}$ on her part, Eve can effectively prepare one of a number of
different ancillary states for Alice and Bob, given by
\begin{equation}
  p_{e} \sigma_{\rA'\rB'|e}
  = \Tr_{\rE} \bigsq{(\id_{\rA'\rB'} \otimes \Pi_{e}) \sigma_{\rA'\rB'\rE}}
\end{equation}
with $p_{e} = \Tr[\Pi_{e} \sigma_{\rE}]$, depending on the outcome $e$ she
obtains. Alice and Bob's part then is the average
\begin{equation}
  \sigma_{\rA'\rB'} = \sum_{e} p_{e} \sigma_{\rA'\rB'|e}
\end{equation}
of these and the joint probabilities of outcomes of POVMs $\{R_{a}\}$ and
$\{S_{b}\}$ they perform are averages
\begin{equation}
  \avg{R_{a} \otimes S_{b}}
  = \sum_{e} p_{e} \avg{R_{a} \otimes S_{b}}_{|e}
\end{equation}
of the corresponding joint probabilities
\begin{equation}
  \avg{R_{a} \otimes S_{b}}_{|e}
  = \Tr \bigsq{(R_{a} \otimes S_{b}) \,
    \psi_{\theta} \otimes \sigma_{\rA'\rB'|e}}
\end{equation}
conditioned on $e$. As we pointed out in the main text, with such a strategy
an eavesdropper can improve her chance of correctly guessing Alice and Bob's
joint outcome if they both use four-outcome POVMs.

Now let us suppose that Alice uses a four-outcome measurement $\{R_{a}\}$ and
Bob a measurement $\{S_{b}\}$ with three outcomes or less. By checking their
correlations with the other measurements in the Bell test, Alice and Bob
infer that the elements of their POVMs are of the forms
\begin{IEEEeqnarray}{rCl}
  R_{a} &=& \alpha_{a} \otimes A'_{+} + \alpha^{*}_{a} \otimes A'_{-}
  \nonumber \\
  &&+\> \trans{\alpha_{a}}{\alpha^{*}_{a}} \otimes K'_{a}
  + \trans{\alpha^{*}_{a}}{\alpha_{a}} \otimes {K'_{a}}^{\dagger} \,,
\end{IEEEeqnarray}
and
\begin{equation}
  S_{b} = \beta_{b} \otimes B'_{+} + \beta^{*}_{b} \otimes B'_{-}
\end{equation}
for some reference extremal qubit POVMs $\{\alpha_{a}\}$ and
$\{\beta_{b}\}$. The joint probability for an underlying state
$\psi_{\theta} \otimes \sigma_{\rA'\rB'|e}$ expands to an expression,
\begin{IEEEeqnarray}{rCl}
  \avg{R_{a} \otimes S_{b}}_{|e}
  &=& \avg{\alpha_{a} \otimes \beta_{b}}_{\psi_{\theta}}
  \avg{A'_{+} \otimes B'_{+}}_{\sigma_{\rA'\rB'|e}} \nonumber \\
  &&+\> \avg{\alpha^{*}_{a} \otimes \beta_{b}}_{\psi_{\theta}}
  \avg{A'_{-} \otimes B'_{+}}_{\sigma_{\rA'\rB'|e}} \nonumber \\
  &&+\> \bavg{
    \trans{\alpha_{a}}{\alpha^{*}_{a}} \otimes \beta_{b}}_{\psi_{\theta}}
  \avg{K'_{a} \otimes B'_{+}}_{\sigma_{\rA'\rB'|e}} \nonumber \\
  &&+\> \bavg{
    \trans{\alpha^{*}_{a}}{\alpha_{a}} \otimes \beta_{b}}_{\psi_{\theta}}
  \avg{{K'_{a}}^{\dagger} \otimes B'_{+}}_{\sigma_{\rA'\rB'|e}} \nonumber \\
  &&+\> \avg{\alpha_{a} \otimes \beta^{*}_{b}}_{\psi_{\theta}}
  \avg{A'_{+} \otimes B'_{-}}_{\sigma_{\rA'\rB'|e}} \nonumber \\
  &&+\> \avg{\alpha^{*}_{a} \otimes \beta^{*}_{b}}_{\psi_{\theta}}
  \avg{A'_{-} \otimes B'_{-}}_{\sigma_{\rA'\rB'|e}} \nonumber \\
  &&+\> \bavg{
    \trans{\alpha_{a}}{\alpha^{*}_{a}} \otimes \beta^{*}_{b}}_{\psi_{\theta}}
  \avg{K'_{a} \otimes B'_{-}}_{\sigma_{\rA'\rB'|e}} \nonumber \\
  &&+\> \bavg{
    \trans{\alpha^{*}_{a}}{\alpha_{a}} \otimes \beta^{*}_{b}}_{\psi_{\theta}}
  \avg{{K'_{a}}^{\dagger} \otimes B'_{-}}_{\sigma_{\rA'\rB'|e}} \,,
  \nonumber \\*
\end{IEEEeqnarray}
which is quite lengthy but we can use properties of the state and POVMs
derived in the previous two sections to show that most of the terms
vanish. We recall first that $\sigma_{\rA'\rB'}$ cannot be arbitrary and must
be such that $\avg{A' \otimes B'}_{\sigma_{\rA'\rB'}} = 1$, which requires
that
\begin{equation}
  \avg{A' \otimes B'}_{\sigma_{\rA'\rB'|e}} = 1
\end{equation}
for all of Eve's possible outcomes $e$. In terms of the projectors $A'_{\pm}$
and $B'_{\pm}$ this means that
\begin{equation}
  \avg{A'_{+} \otimes B'_{+}}_{\sigma_{\rA' \rB'|e}}
  + \avg{A'_{-} \otimes B'_{-}}_{\sigma_{\rA' \rB'|e}} = 1
\end{equation}
and
\begin{equation}
  \label{eq:ApmBmp_correlation}
  \avg{A'_{+} \otimes B'_{-}}_{\sigma_{\rA' \rB'|e}}
  = \avg{A'_{-} \otimes B'_{+}}_{\sigma_{\rA' \rB'|e}} = 0 \,.
\end{equation}
The second pair \eqref{eq:ApmBmp_correlation} of constraints already means
that the terms involving
$\avg{\alpha^{*}_{a} \otimes \beta_{b}}_{\psi_{\theta}}$ and
$\avg{\alpha_{a} \otimes \beta^{*}_{b}}_{\psi_{\theta}}$
vanish. \eqref{eq:ApmBmp_correlation} furthermore implies the operator
constraints
\begin{IEEEeqnarray}{rCl}
  (A'_{+} \otimes B'_{-}) \sqrt{\sigma_{\rA'\rB'|e}} &=& 0 \,, \\
  (A'_{-} \otimes B'_{+}) \sqrt{\sigma_{\rA'\rB'|e}} &=& 0 \,,
\end{IEEEeqnarray}
while we recall that the $K'_{a}$s satisfy
\begin{equation}
  A'_{+} K'_{a} = K'_{a} A'_{-} = K'_{a} \,.
\end{equation}
These constraints can be used to show, for example,
\begin{IEEEeqnarray}{rCl}
  \avg{K'_{a} \otimes B'_{+}}_{\sigma_{\rA'\rB'|e}}
  &=& \Tr \bigsq{(K'_{a} A'_{-} \otimes B'_{+}) \, \sigma_{\rA'\rB'|e}}
  \nonumber \\
  &=& 0 \,,
\end{IEEEeqnarray}
with similar manipulations implying
\begin{IEEEeqnarray}{rCl}
  \avg{{K'_{a}}^{\dagger} \otimes B'_{+}}_{\sigma_{\rA'\rB'|e}} &=& 0 \,, \\
  \avg{K'_{a} \otimes B'_{-}}_{\sigma_{\rA'\rB'|e}} &=& 0 \,, \\
  \avg{{K'_{a}}^{\dagger} \otimes B'_{-}}_{\sigma_{\rA'\rB'|e}} &=& 0 \,.
\end{IEEEeqnarray}
In the joint probabilities we are thus left with only
\begin{IEEEeqnarray}{rCl}
  \avg{R_{a} \otimes S_{b}}_{|e}
  &=& \avg{\alpha_{a} \otimes \beta_{b}}_{\psi_{\theta}}
  \avg{A'_{+} \otimes B'_{+}}_{\sigma_{\rA'\rB'|e}} \nonumber \\
  &&+\> \avg{\alpha^{*}_{a} \otimes \beta^{*}_{b}}_{\psi_{\theta}}
  \avg{A'_{-} \otimes B'_{-}}_{\sigma_{\rA'\rB'|e}} \,, \IEEEeqnarraynumspace
\end{IEEEeqnarray}
which simplify to
\begin{equation}
  \avg{R_{a} \otimes S_{b}}_{|e}
  = \avg{\alpha_{a} \otimes \beta_{b}}_{\psi_{\theta}}
\end{equation}
because
\begin{equation}
  \avg{\alpha^{*}_{a} \otimes \beta^{*}_{b}}_{\psi_{\theta}}
  = \avg{\alpha_{a} \otimes \beta_{b}}_{\psi^{*}_{\theta}}
  = \avg{\alpha_{a} \otimes \beta_{b}}_{\psi_{\theta}}
\end{equation}
due to the state $\ket{\psi_{\theta}}$ being real.

We conclude that the joint probability of Alice and Bob obtaining outcomes
$a$ and $b$ using the POVMs $\{R_{a}\}$ and $\{S_{b}\}$, even conditioned on
the outcome of a measurement by Eve, is identical to the probability of
obtaining the same outcomes using the reference qubit POVMs $\{\alpha_{a}\}$
and $\{\beta_{b}\}$ on $\ket{\psi_{\theta}}$. This considerably simplifies
the problem of determining how much randomness we can generate using POVMs
provided that a four-outcome POVM is used only on one side.

\subsection{Two bits of local randomness}

We suppose here that Alice wishes to extract two bits of randomness in the
context of our Bell test. Following the analysis of the previous subsection,
we need only confirm that there exist extremal qubit POVMs that produce a
uniformly random result from the ideal partially entangled state
$\ket{\psi_{\theta}}$.

In the ideal qubit setting, Alice's marginal state is
\begin{equation}
  \psi_{\theta\rA} = \tfrac{1}{2} \bigro{\id + \cos(\theta) \sz}
\end{equation}
and we are looking to construct a four-outcome POVM $\{\alpha_{a}\}$ where
the POVM elements $\alpha_{a}$ are of rank one, are linearly independent, and
yield the correct probabilities
\begin{equation}
  \Tr \bigsq{\alpha_{a} \psi_{\theta\rA}} = \frac{1}{4} \,.
\end{equation}
Such POVMs can be constructed by adjusting a tetrahedral POVM, which gives
the correct amount of randomness in the special case of the maximally-mixed
state. A specific example that works is to take
\begin{equation}
  \alpha_{1} = \frac{\lambda_{1}}{2} \bigro{\id + \sz}
\end{equation}
and
\begin{IEEEeqnarray}{rl}
  \label{eq:alphaa_example}
  \alpha_{a} = \frac{\lambda_{a}}{2}
  \Bigro{&\id + \cos(\gamma) \sz \nonumber \\
    &+\> \sin(\gamma) \bigro{\cos(\delta_{a}) \sx
      + \sin(\delta_{a}) \sy}} \IEEEeqnarraynumspace
\end{IEEEeqnarray}
with weights
\begin{IEEEeqnarray}{rCl.l}
  \lambda_{1} &=& \frac{1}{2 + 2 \cos(\theta)} \,, & \\
  \lambda_{a} &=& \frac{3 + 4 \cos(\theta)}{6 + 6 \cos(\theta)} &
  \text{ for } a = 2, 3, 4
\end{IEEEeqnarray}
and, in \eqref{eq:alphaa_example}, an angle $\gamma$ such that
\begin{equation}
  \cos(\gamma) = -\frac{1}{3 + 4 \cos(\theta)}
\end{equation}
and for example $\delta_{2}, \delta_{3}, \delta_{4} = 0, 2\pi/3, 4\pi/3$.

\subsection{$\log_{2}(12)$ bits of global randomness}

In this second variant, we suppose Alice and Bob both use POVMs to generate
randomness from their joint outcomes. If one of the parties, say, Bob, uses a
POVM with three outcomes then it is in principle possible to generate up to
$\log_{2}(12)$ bits of randomness. In the ideal qubit setting, at least one
way to get arbitrarily close to attaining this is for Bob to perform a
modified version of a Mercedes-Benz POVM in the $\sx$-$\sz$ plane,
\begin{IEEEeqnarray}{rCl}
  \beta_{1} &=& \frac{\lambda_{1}}{2} \Bigro{\id + \sz} \,, \\
  \beta_{2} &=& \frac{\lambda_{2}}{2} \Bigro{
    \id - \mu \sz + \sqrt{1 - \mu^{2}} \, \sx} \,, \\
  \beta_{3} &=& \frac{\lambda_{3}}{2} \Bigro{
    \id - \mu \sz - \sqrt{1 - \mu^{2}} \, \sx}
\end{IEEEeqnarray}
for parameters related to $\theta$ by
\begin{IEEEeqnarray}{rCl}
  \lambda_{1} &=& \frac{2}{3 + 3 \cos(\theta)} \,, \\
  \lambda_{2} = \lambda_{3}
  &=& \frac{2 + 3 \cos(\theta)}{3 + 3 \cos(\theta)} \,, \\
  \mu &=& \frac{1}{2 + 3 \cos(\theta)} \,,
\end{IEEEeqnarray}
while Alice performs a measurement
\begin{IEEEeqnarray}{rCl}
  \alpha_{1} &=& \frac{1}{4} \Bigro{
    \id + \sqrt{1 - \varepsilon^{2}} \, \sy + \varepsilon \sz} \,, \\
  \alpha_{2} &=& \frac{1}{4} \Bigro{
    \id + \sqrt{1 - \varepsilon^{2}} \, \sy - \varepsilon \sz} \,, \\
  \alpha_{3} &=& \frac{1}{4} \Bigro{
    \id - \sqrt{1 - \varepsilon^{2}} \, \sy + \varepsilon \sx} \,, \\
  \alpha_{4} &=& \frac{1}{4} \Bigro{
    \id - \sqrt{1 - \varepsilon^{2}} \, \sy - \varepsilon \sx}
\end{IEEEeqnarray}
whose elements deviate only a small amount from the $\sy$ axis. Importantly,
Alice's measurement is extremal for any nonzero $\varepsilon$. For small
$\varepsilon$, these measurements give joint probabilities
\begin{equation}
  \avg{\alpha_{a} \otimes \beta_{b}}_{\psi_{\theta}}
  \approx \frac{1}{12} \,,
\end{equation}
with equality in the limit $\varepsilon \to 0$.

\end{document}